\documentclass[a4paper, 12pt]{elsarticle}

\usepackage{graphics}
\usepackage{graphicx}
\usepackage{psfig}
\usepackage{rotating}
\usepackage[absolute]{textpos}
\usepackage{color}
\usepackage{lineno}
\usepackage[algo2e]{algorithm2e} 
\usepackage[utf8]{inputenc}
\usepackage{amsmath}

\usepackage{amsfonts}
\usepackage{amssymb}
\usepackage{amsthm}
\usepackage{mathrsfs}
\usepackage{latexsym}
\usepackage{multirow}
\usepackage{rotating}
\usepackage{caption}
\usepackage{graphicx}
\usepackage{float}
\usepackage{array}
\usepackage{subfigure}
\usepackage{tabularx}
\usepackage{graphicx,psfig}
\usepackage{footnote}
\usepackage[titletoc,toc,title]{appendix}
\usepackage{colortbl}
\usepackage[dvipsnames]{xcolor}
\usepackage{hyperref}
\usepackage{natbib}
\usepackage{amssymb}
\usepackage{soul}
\usepackage{algorithmic}
\usepackage{algorithm}
\usepackage{epsfig}
\usepackage{relsize}
\usepackage{color}
\usepackage{mathtools}
\usepackage{graphicx,booktabs}
\usepackage{multirow}
\usepackage[utf8]{inputenc}
\usepackage[english]{babel}

\begin{document}

\title{An ensemble neural network approach to forecast Dengue outbreak based on climatic condition}

\author{Madhurima Panja\textsuperscript{1,}
\footnote[0]{\textit{Equal Contributions}}, 
Tanujit Chakraborty\textsuperscript{1, 0,}
\footnote[2]{\textit{Corresponding author}: \textit{Mail}: tanujit.chakraborty@sorbonne.ae},
Sk Shahid Nadim\textsuperscript{3},
Indrajit Ghosh\textsuperscript{4},
Uttam Kumar\textsuperscript{1},
Nan Liu\textsuperscript{5}\\
{\scriptsize \textsuperscript{1} Spatial Computing Laboratory, Center for Data Sciences, International Institute of Information Technology Bangalore, India.}\\
{\scriptsize \textsuperscript{2} Department of Science and Engineering, Sorbonne University Abu Dhabi and Sorbonne Center for Artificial Intelligence.} \\
{\scriptsize \textsuperscript{3} Odum School of Ecology, University of Georgia, Athens, Georgia, USA.}\\
{\scriptsize \textsuperscript{4} Department of Epidemiology and Biostatistics, University of Georgia, Athens, Georgia, USA.}\\
{\scriptsize \textsuperscript{5} Duke-NUS Medical School, National University of Singapore, Singapore, Singapore.}}

\begin{abstract}
Dengue fever is a virulent disease spreading over 100 tropical and subtropical countries in Africa, the Americas, and Asia. This arboviral disease affects around 400 million people globally, severely distressing the healthcare systems. The unavailability of a specific drug and ready-to-use vaccine makes the situation worse. Hence, policymakers must rely on early warning systems to control intervention-related decisions. Forecasts routinely provide critical information for dangerous epidemic events. However, the available forecasting models (e.g., weather-driven mechanistic, statistical time series, and machine learning models) lack a clear understanding of different components to improve prediction accuracy and often provide unstable and unreliable forecasts. This study proposes an ensemble wavelet neural network with exogenous factor(s) (XEWNet) model that can produce reliable estimates for dengue outbreak prediction for three geographical regions, namely San Juan, Iquitos, and Ahmedabad. The proposed XEWNet model is flexible and can easily incorporate exogenous climate variable(s) confirmed by statistical causality tests in its scalable framework. The proposed model is an integrated approach that uses wavelet transformation into an ensemble neural network framework that helps in generating more reliable long-term forecasts. The proposed XEWNet allows complex non-linear relationships between the dengue incidence cases and rainfall; however, mathematically interpretable, fast in execution, and easily comprehensible. The proposal's competitiveness is measured using computational experiments based on various statistical metrics and several statistical comparison tests. In comparison with statistical, machine learning, and deep learning methods, our proposed XEWNet performs better in 75\% of the cases for short-term and long-term forecasting of dengue incidence. 
\end{abstract}
%\end{frontmatter}
\maketitle
\section{Introduction} \label{Introduction}
Dengue, a fatal viral disease, is a major public health concern in many tropical and subtropical regions of the globe \cite{bhatt2013global}. According to a recent article \cite{zeng2021global}, the global number of dengue fever cases has increased from 23 million in 1990 to 104 million in 2017. While this infectious disease mainly affects under-developed and developing countries, ongoing climate change and globalization contribute significantly to increasing dengue transmission in previously dengue-free nations. The spectrum of dengue infection ranges from febrile disease, including high fever, retro-ocular pain, muscle, and joint pain, skin rashes, and headaches, to severe manifestations of hemorrhagic fever and shock syndrome \cite{guzman2015dengue}. Due to the unavailability of specific antiviral treatment or ready-to-use licensed vaccines, public health officials primarily rely on supportive treatments and insecticide-based vector control interventions to limit the spread of dengue infection \cite{murphy2011immune}. While the disease is endemic in several regions of Africa, the western Pacific, and Southeast Asia, these areas recurrently experience a high risk of infection when the mosquito population surges near the human habitat. Climatic factors such as rainfall, temperature, and relative humidity have been found to influence the rapid spread of the dengue virus \cite{polwiang2020time}. In such cases, the Aedes mosquitoes undergo shorter times of breeding and maturity period, resulting in higher population growth \cite{watts1987effect}. Moreover, owing to the shrinkage of the incubation period in the Aedes vectors, their ability to infect humans with the dengue virus increases rapidly \cite{focks2000transmission}. Recent studies \cite{wu2007weather,yang2009assessing} have shown that meteorological conditions significantly influence the size of dengue dispersion through their impacts on the life cycle developments, biting rates, infective, and survivability rates of vectors. Similar results evaluating the effect of heavy rainfall on rapid growth in vector populations have been discussed in \cite{hii2009climate}.

With the increasing spread of dengue infection due to climatic changes and unplanned urbanization, there has been an unprecedented burden on the healthcare infrastructure. Studies have identified that early detection and treatment of severe cases can significantly reduce the risk of health complications and mortality \cite{degallier2010toward}. Moreover, to plan and allocate resources, accurate predictions of infected individuals, or incidence, are crucial. For instance, knowing the expected number of cases and when they will occur enables planning through education and awareness campaigns, resource reallocation to high-risk regions, or medical retraining to detect symptoms and treat them correctly before peak transmission \cite{thomson2008seasonal}. Due to this, public health authorities have to rely on early warning systems to disseminate available resources in the dengue-endemic areas optimally. The effectiveness of accurate dengue forecast in designing pre-epidemic preparedness strategies has attracted many researchers to attempt dengue incidence and outbreak prediction with different degrees of success \cite{gharbi2011time,buczak2018ensemble,lauer2018prospective}. These efforts were put into establishing an accurate early warning system to inform healthcare officials about an upcoming outbreak. The type of models used to predict dengue cases varied diversely. Previous works have used mechanistic models \cite{racloz2012surveillance}, some others rely on statistical models \cite{johnson2018phenomenological}, data-driven machine learning techniques \cite{guo2017developing}, and few others have used ensemble forecasting systems \cite{yamana2016superensemble,deb2022ensemble}. The use of mechanistic models based on deterministic differential equations with detailed biology of dengue transmission has often failed to generate reliable forecasts because of the data needed to parameterize the models. Another study by \cite{rangarajan2019forecasting} incorporated data from social-media activity and internet searches to forecast dengue incidence.  To this end, forecasters have primarily relied on the dengue cases and environmental factors and used statistical and machine learning models with a higher degree of success than the mechanistic models \cite{chakraborty2019forecasting, yamana2016superensemble, deb2022ensemble}. 

Numerous statistical time series and machine learning models have been developed for solving the dengue forecasting problem. Time series approaches, either using covariates or historical incidence data alone, have been widely used in several studies \cite{luz2008time, promprou2006forecasting}. Even among the statistical methodologies, the auto-regressive integrated moving average (ARIMA) model is the most suitable technique for lagged historical incidence data observations to generate one-step and multi-step ahead forecasts. While the statistical frameworks primarily focus on parametric methods to model the complexities of dengue incidence cases, recent studies manifest the use of advanced machine learning and deep learning approaches to generate reliable epidemic forecasts \cite{guo2017developing,bhattacharyya2022stochastic}. Another emerging technique in this regard is the ensemble framework that enhances the forecasting efficiency of individual models. In \cite{deb2022ensemble}, the authors have proposed an ensemble approach combining negative binomial regression, ARIMA, and generalized linear auto-regressive moving average model through a vector auto-regressive structure to generate early forecasts for the dengue outbreak. On the other hand, multiple studies have manifested that climate variation is one of the critical drivers of dengue outbreaks. In \cite{johansson2016evaluating}, the researchers have utilized the seasonal ARIMA model to study the interrelationship of weather data and dengue incidence series in Mexico. In addition, recent studies have recommended using a generalized additive model with lags in the number of cases and meteorological variables to capture the seasonal patterns of dengue outbreak \cite{baquero2018dengue}. Moreover, the use of a superensemble framework in \cite{colon2021probabilistic} has increased accuracy in the dengue forecast by combining Earth observational data, seasonal climate forecasts, and state-of-the-art models. Moreover, recent studies have utilized artificial neural networks \cite{zhao2020machine}, auto-regressive neural networks (ARNN) \cite{baquero2018dengue}, and hybrid ARIMA-ARNN model \cite{chakraborty2019forecasting} for univariate forecasting of dengue surveillance data for severely affected regions. Convolutional neural network (CNN), Transformer, long short-term memory (LSTM), and attention-enhanced LSTM (LSTM-ATT) models were compared with traditional machine learning models on weather-based dengue fever forecasting for Vietnam in \cite{nguyen2022deep}.

Albeit the vast application of statistical, machine learning, deep learning, and ensemble framework in dengue forecasting, these conventional models suffer from some significant drawbacks. For instance, the statistical methods, especially the linear ARIMA and ETS methods, are insufficient to model the non-linear complexities in real-world dengue datasets. Although advanced machine learning and deep learning frameworks can handle the irregular behavior of the epidemic series using canonical architectures, their forecasting performance severely diminishes for long-term prediction problems. Moreover, these data-driven methods suffer from the problem of overfitting, i.e., it learns the patterns and the noise in the training series to such an extent that it negatively impacts the forecasting performance for an unseen dataset. In the case of ensemble frameworks, although the forecast accuracy enhances w.r.t the constituent models, the appropriate selection of models and their weights, often termed `forecast combination puzzle', pose a significant challenge to its universal success. To overcome these drawbacks, wavelet transformation has been considered a proper mathematical technique for decades \cite{walden2001wavelet, percival1997analysis}. The efficacy of the wavelet decomposition to utilize high and low-frequency filters for appropriate signal extraction from noisy time series datasets has led to its popularity in forecasting literature \cite{WARIMA,zhang2017application,fay2007wavelet}. Several studies have utilized the maximal overlapping version of discrete wavelet transformation (MODWT) to model the non-stationary behavior of time series datasets from various domains, namely energy \cite{anjoy2019comparative}, meteorology \cite{nury2017comparative}, and natural calamities \cite{nanda2016wavelet} to name a few.
The wavelet-based forecasting technique has also gained popularity in epidemic forecasting. In \cite{chakraborty2020real}, the researchers have proposed a hybrid ARIMA and wavelet forecasting approach to generate real-time forecasts for the novel Covid-19 pandemic. Recently, an ensemble wavelet neural network (EWNet) architecture for modeling univariate epidemic datasets has been proposed \cite{panja2022epicasting}. This study utilizes a MODWT decomposition technique on the historical incidence data and generates epidemic forecasts by modeling multiple transformed series using auto-regressive neural networks in an ensemble framework. The EWNet model has been shown to improve the forecast accuracy for various epidemics compared to state-of-the-art forecasters. However, this framework is incompetent to handle the causal relationship between epidemic incidence and other climatic factors. Our study establishes that rainfall is a crucial driver of dengue ecology by utilizing robust causality tests; hence, it should be considered in the dengue forecasting framework. Our proposed model extends the EWNet framework to handle the effect of the causal covariate in its architecture. The proposed XEWNet utilizes an ensemble framework to model the wavelet-transformed dengue incidence series and rainfall data using auto-regressive neural networks. Experimental results suggest significant improvement of our proposed model over the EWNet architecture and other benchmark forecasters for short-term and long-term dengue forecasting in the San Juan, Iquitos, and Ahmedabad regions. We also provide the appropriate choices of hyperparameters of the proposed architecture for different horizons in the geographical regions. The statistical tests also validate the robustness of our proposed model. Furthermore, we establish that the current formulation of the XEWNet framework is more suitable as an early warning system for dengue-prone regions than the state-of-the-art forecasters. Moreover, the proposed model generates the most reliable dengue forecasts with one year lead time; hence this can be utilized by public health officials in designing effective dengue combat policies. 

The remaining sections of the paper are organized as follows. First, in Section \ref{Mat_Method}, we discuss the datasets considered in this study, understand the causal relationship between rainfall and dengue incidence datasets, and elaborate on the development of the ensemble model. Next, a detailed experimental evaluation of the proposed XEWNet framework for the short-term and long-term forecast is discussed in Section \ref{Exp_results}. Then, in Section \ref{stat_signif}, we manifest the robustness of our proposed model using statistical tests. Eventually, in Section \ref{discussion}, we conclude the paper with discussions about the results, policy recommendations, and future scope of work. 

\section{Materials and Methods}\label{Mat_Method}

This study is focused on the weekly laboratory-confirmed dengue incidence cases in three tropical and sub-tropical regions: San Juan, Iquitos, and Ahmedabad. To assess the impact of climatic conditions on dengue outbreaks, we consider the data on the weekly distribution of rainfall for these cities. We retrieved the dengue incidence data for San Juan and Iquitos from the National Oceanic and Atmospheric Administration (NOAA) Repository\footnote{\url{https://dengueforecasting.noaa.gov/}}, which was initially provided by the collaboration of the U.S. government, public-sector agencies of Peru, and the U.S. universities. On the other hand, the precipitation data for these regions are accumulated by the U.S. climatology agency and the NOAA. The data on dengue incidence (per 10000 population) and rainfall in the Ahmedabad region has been collected from \cite{enduri2017estimation}. In the case of San Juan, the weekly data is considered from the beginning of 1991 till the end of 2012. For Iquitos and Ahmedabad, the period ranging from 2002 to 2012 and 2005 to 2012, respectively, have been contemplated in this study. In the adjoining section, we describe the global characteristics of these datasets in detail. The data and codes used in this study are made publicly accessible at \url{https://github.com/mad-stat/XEWNet} for the reproducibility of this work.

\begin{figure}
    \centering
    \includegraphics[width=0.325\textwidth]{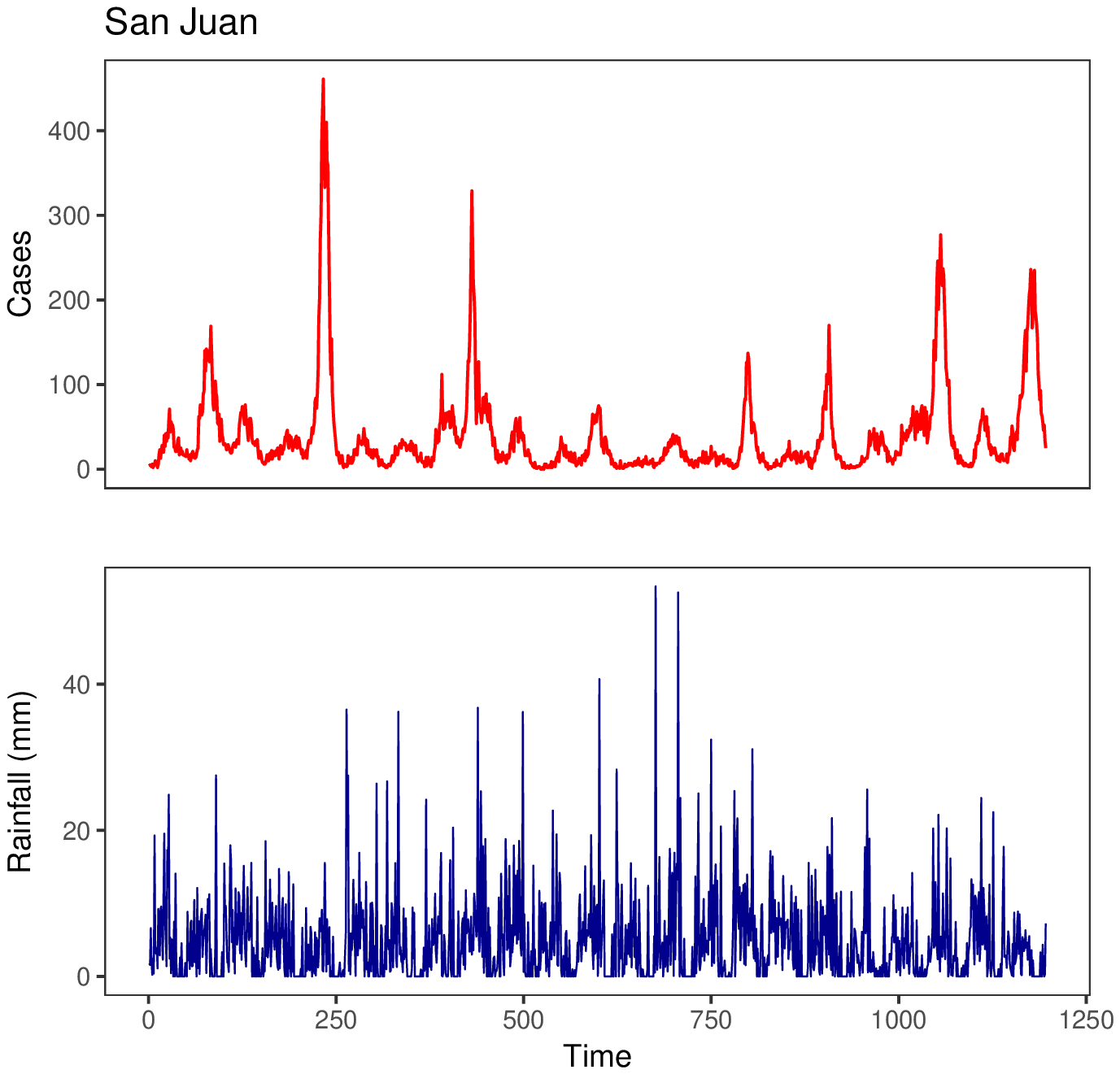}
    \includegraphics[width=0.325\textwidth]{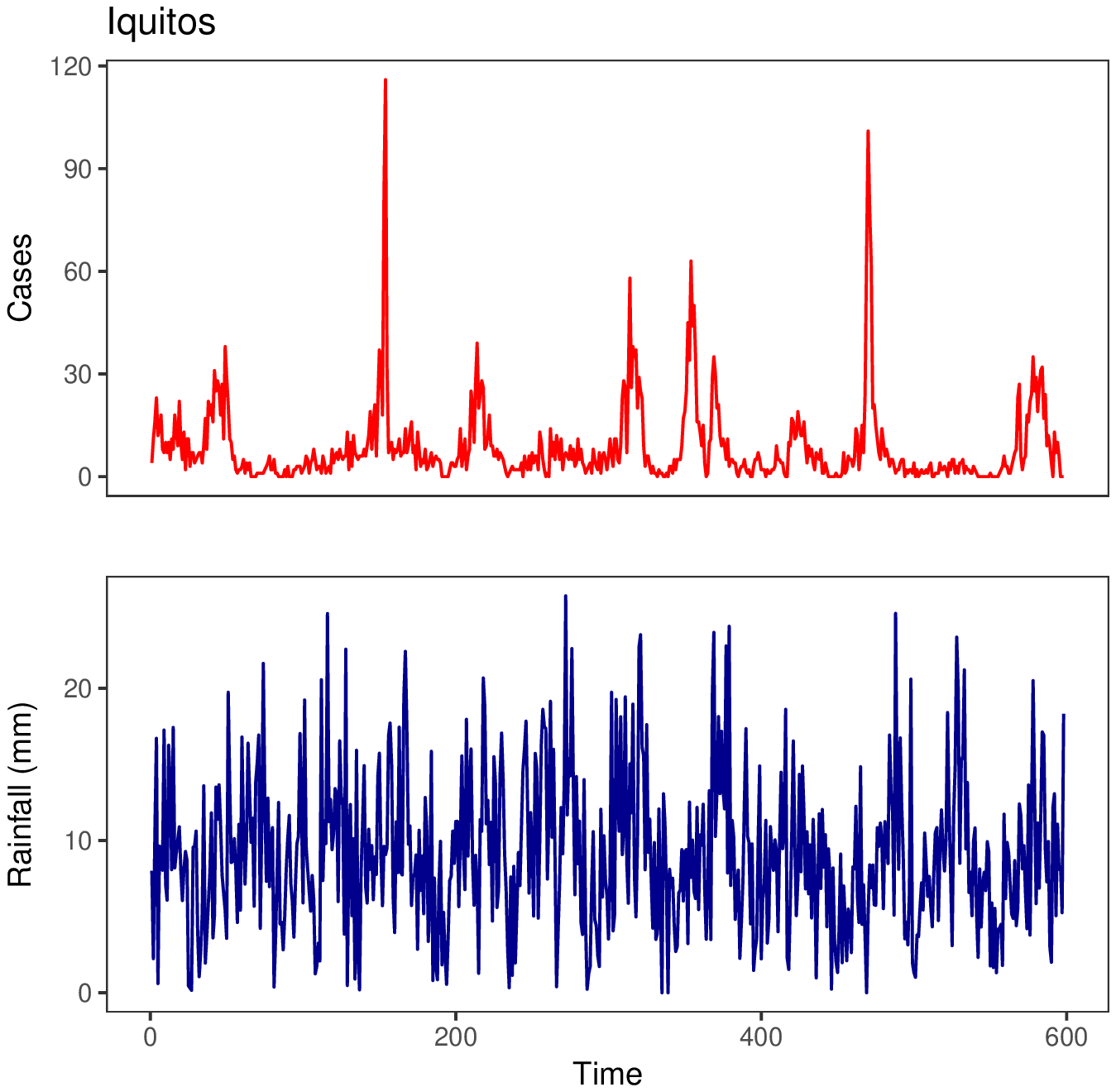}
    \includegraphics[width=0.325\textwidth]{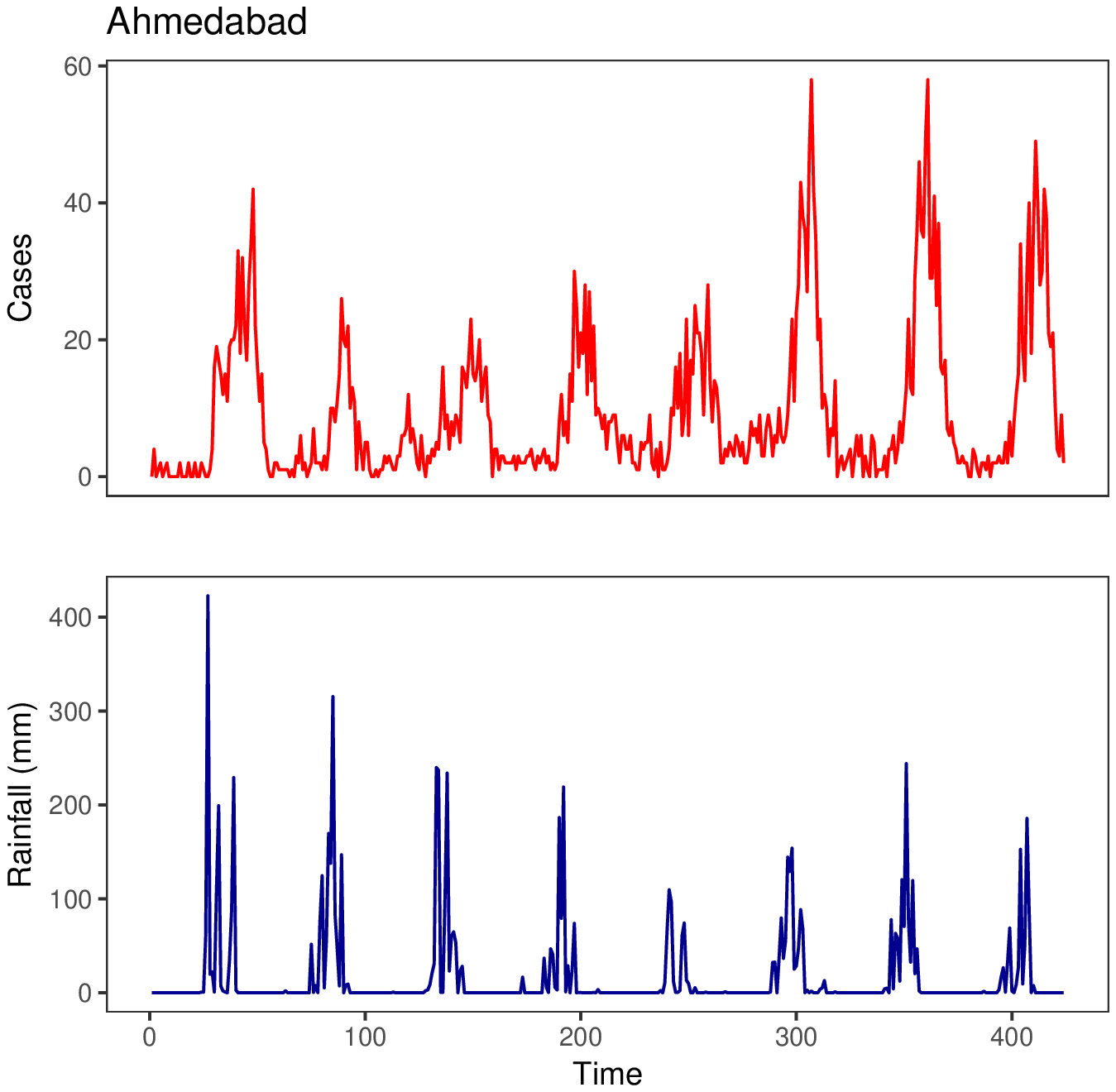}
    \caption{Dengue cases (red) and rainfall (blue) data of San Juan, Iquitos, and Ahmedabad (incidence per 10000 population).}
    \label{fig:data_rainfall}
\end{figure}

\subsection{Dengue Surveillance Data}
This study uses three multivariate time series datasets (weekly) to generate a short-term (26 weeks) and long-term (52 weeks) forecast of dengue incidence in the San Juan, Iquitos, and Ahmedabad regions. Previous studies have used several traditional time series models and modern deep learning architectures to forecast the dengue outbreak in San Juan and Iquitos cities; for details, see \cite{buczak2018ensemble, johnson2018phenomenological, johansson2019open, deb2022ensemble}. In this study, we first intend to test the hypothesis that rainfall plays a crucial role in dengue transmission. For doing so, we mainly focus on regions within the subtropical belt and receiving moderate to heavy rainfall. For instance, San Juan situated along the southern side of the Atlantic Ocean across the northeastern coast of Puerto Rico experiences a tropical monsoon climate with frequent rainfall throughout the year. The available data shows that the average rainfall in this region is 4.78 mm, ranging from 0 to 53.39 mm from 1991 to 2012. Iquitos, the northeastern Peruvian city, is situated south of the Amazon River. This region experiences a hot and humid equatorial climate with wet summers. Nevertheless, the town encountered constant precipitation, with average weekly rainfall of 9.09 mm, ranging from 0 to 26.07 mm from 2002 to 2012. On the other hand, the north-western metropolitan city of India, Ahmedabad, is located along the banks of the Sabarmati River. A tropical monsoon climate prevails in this city throughout the year, with moderate precipitation during the monsoon season. This region encountered rainfall ranging from 0 to 422.62 mm from 2005 to 2012. A thematic plot representing rainfall distribution and dengue incidence cases for the studied areas is presented in Fig. \ref{fig:data_rainfall}. In the subsequent section, we provide a detailed description of the global characteristics of the dengue datasets considered in this study that will be useful for building a dengue forecasting system for its probable usage in national public health systems. 

\begin{table} 
    \centering
    \caption{Dengue incidence datasets and their corresponding ACF, PACF plots for three regions}\label{table_acf_pacf} \vspace{1cm}
    \begin{tabular}{ | c | p{4cm} | p{4cm} | p{4cm} |}
        \hline
        Region & Data & ACF plot & PACF plot  \\ \hline
        San Juan
        &
        \begin{minipage}{.30\textwidth}
            \includegraphics[width=40mm, height=30mm]{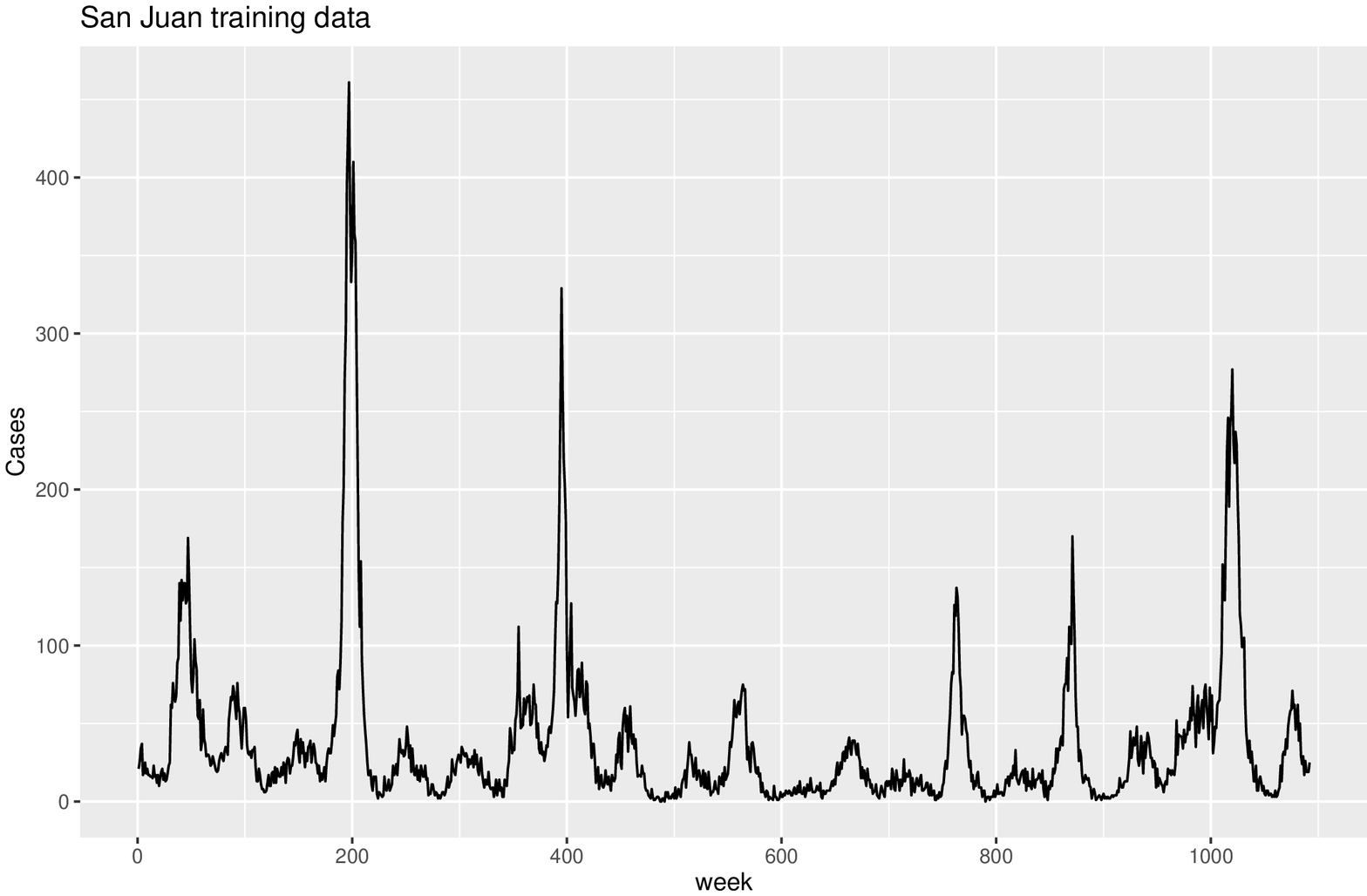}
        \end{minipage}
        &
        \begin{minipage}{.30\textwidth}
            \includegraphics[width=40mm, height=30mm]{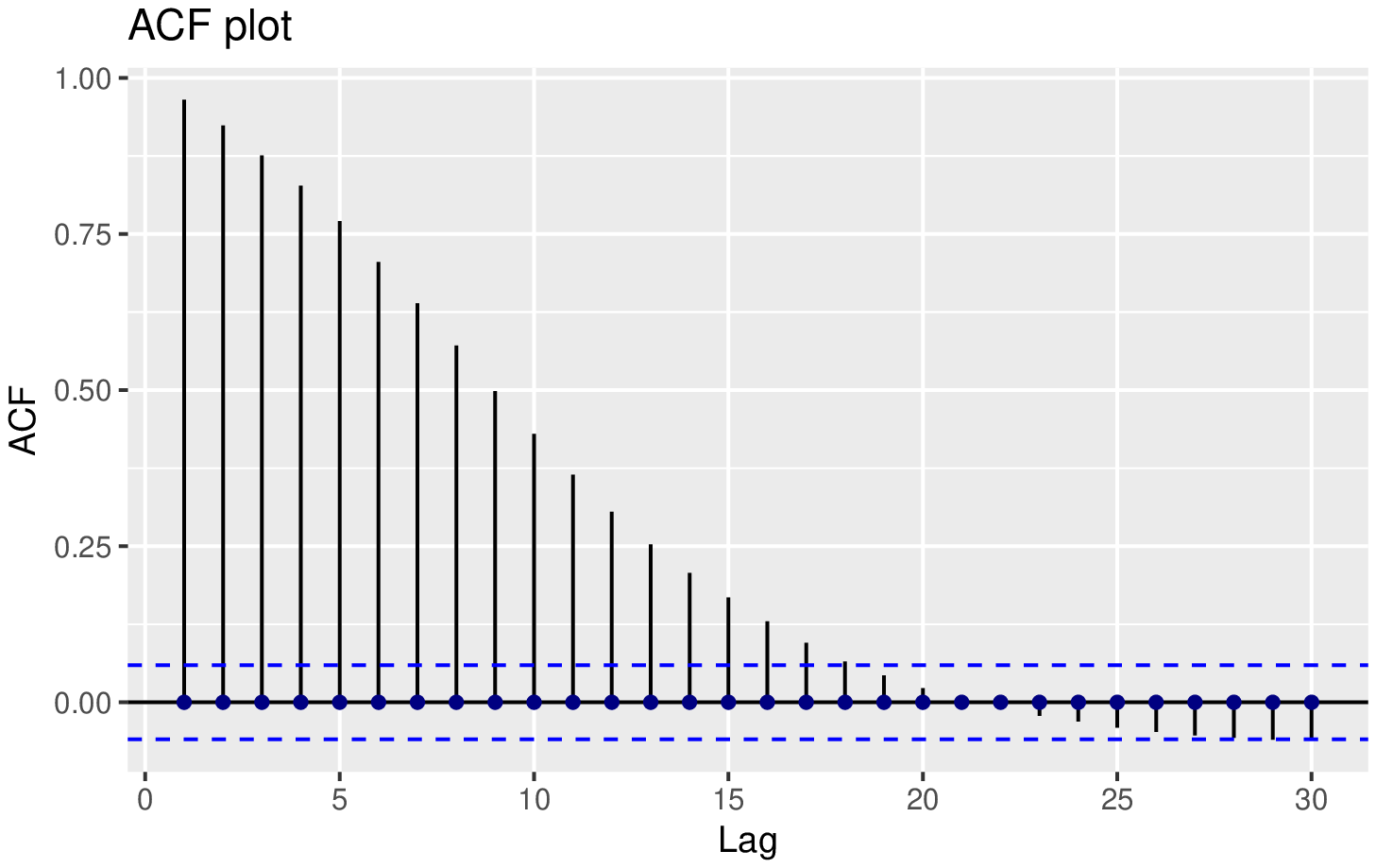}
        \end{minipage}
        &
        \begin{minipage}{.30\textwidth}
            \includegraphics[width=40mm, height=30mm]{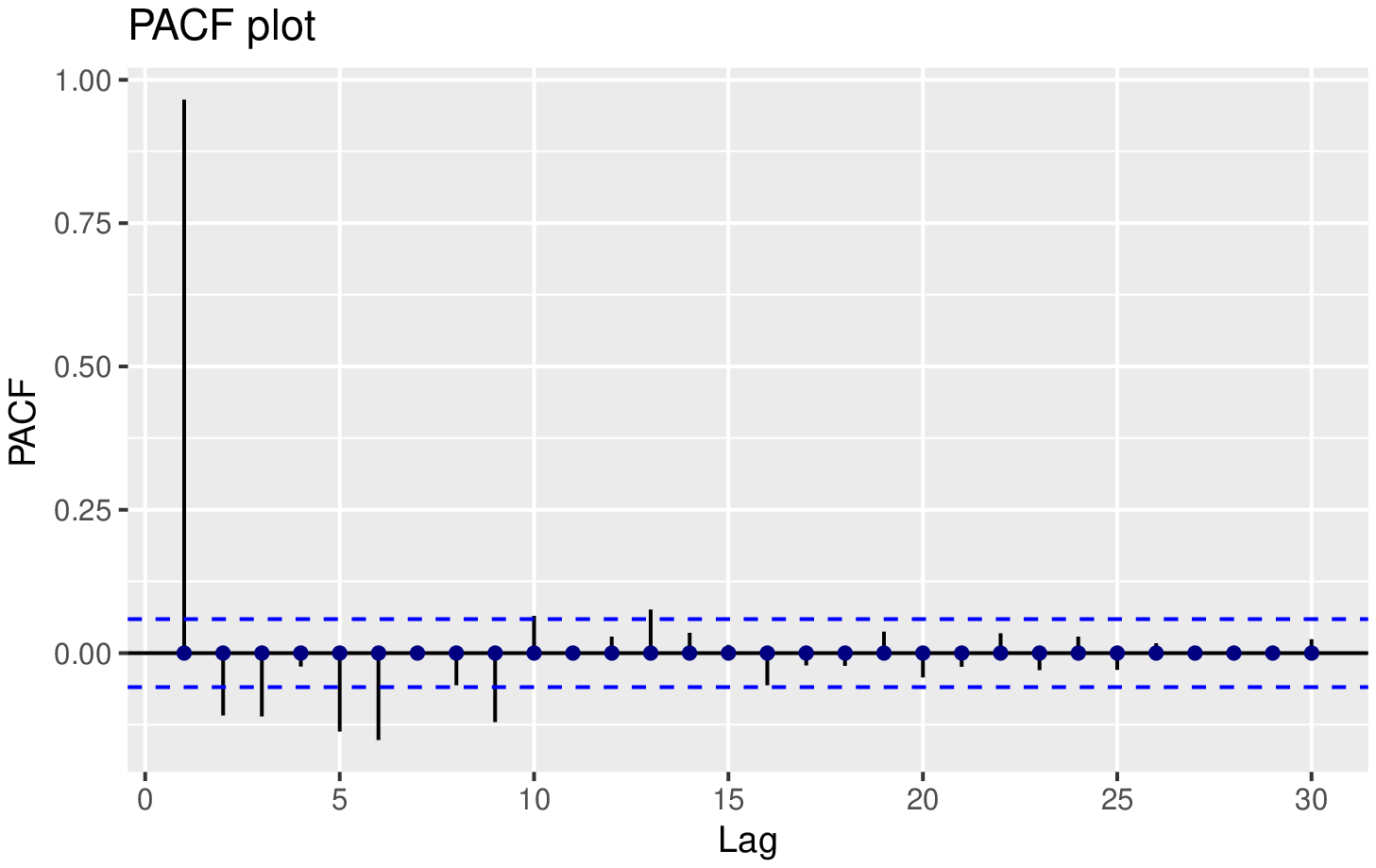}
        \end{minipage} \\ \hline
        Iquitos
        &
        \begin{minipage}{.3\textwidth}
            \includegraphics[width=40mm, height=30mm]{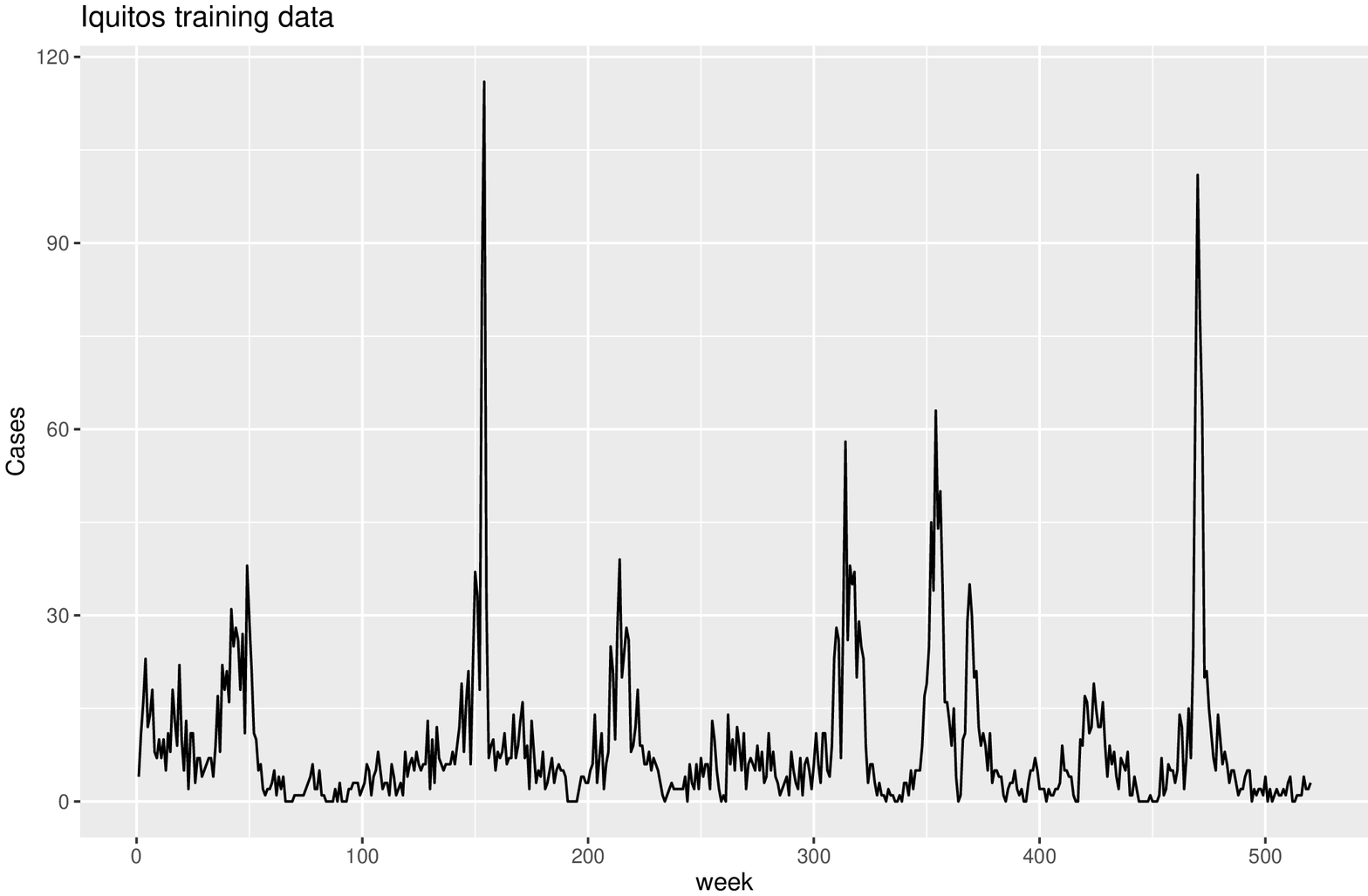}
        \end{minipage}
        &
        \begin{minipage}{.3\textwidth}
            \includegraphics[width=40mm, height=30mm]{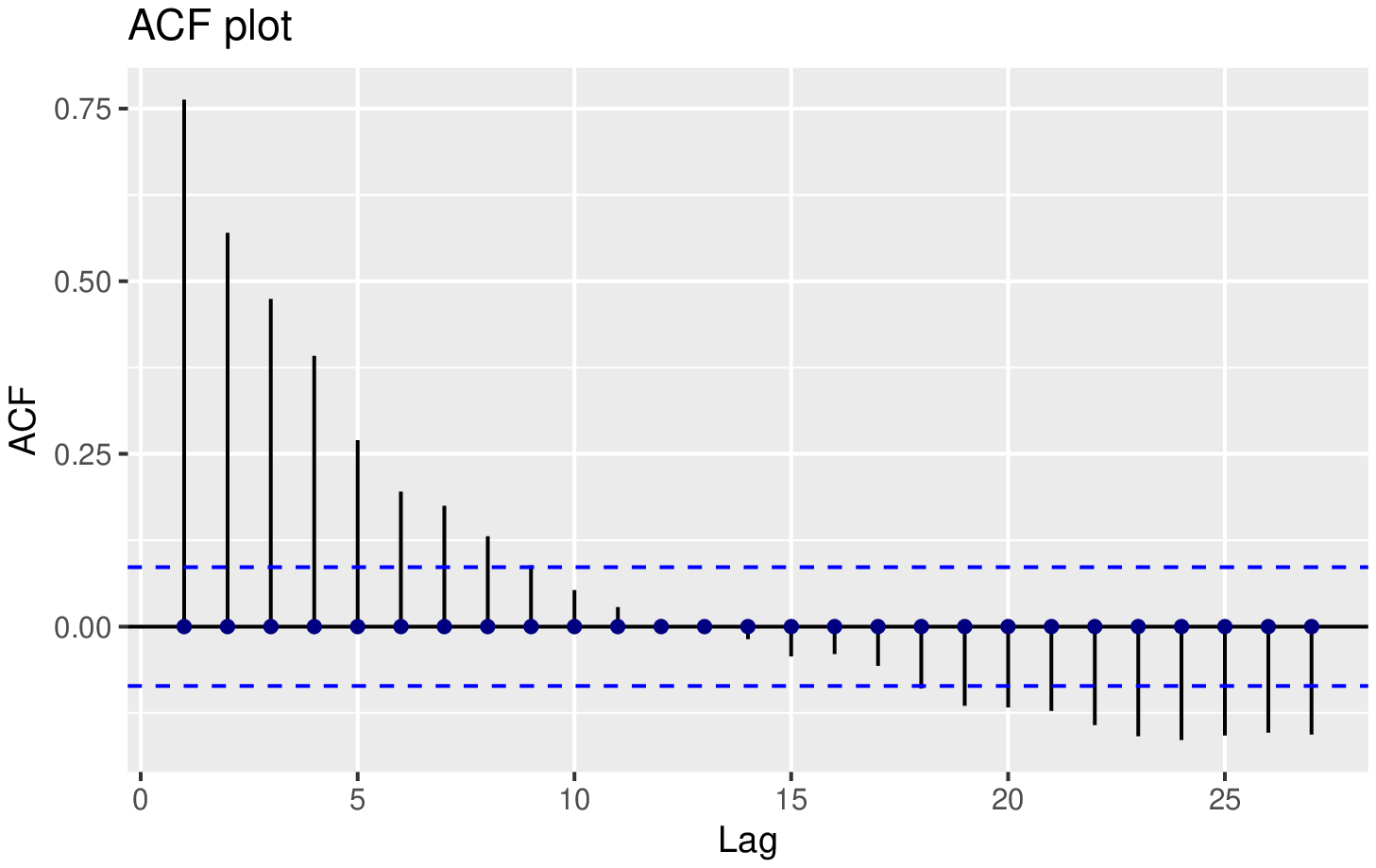}
        \end{minipage}
        &
        \begin{minipage}{.3\textwidth}
            \includegraphics[width=40mm, height=30mm]{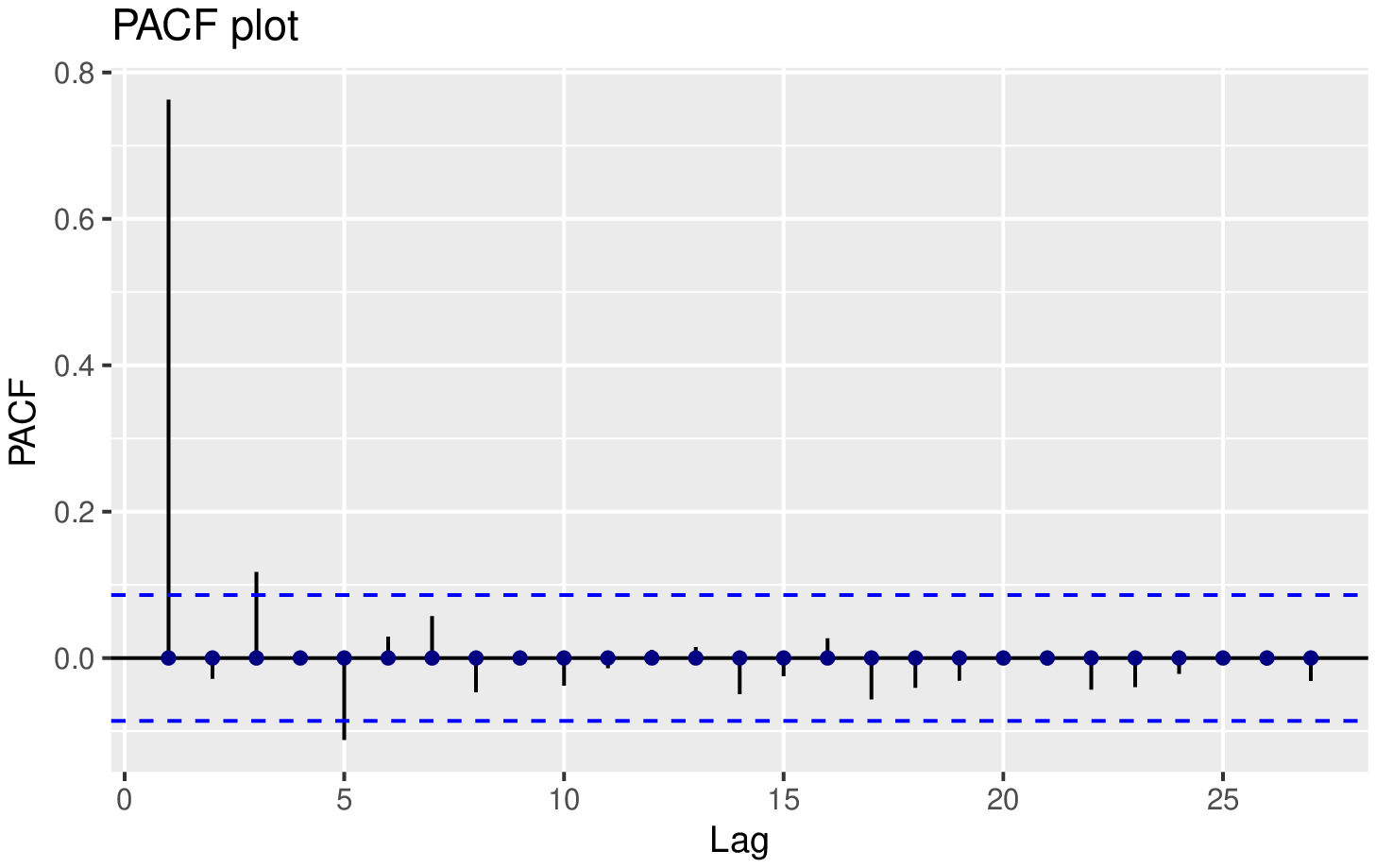}
        \end{minipage}
        \\ \hline
        Ahmedabad
        &
        \begin{minipage}{.3\textwidth}
            \includegraphics[width=40mm, height=30mm]{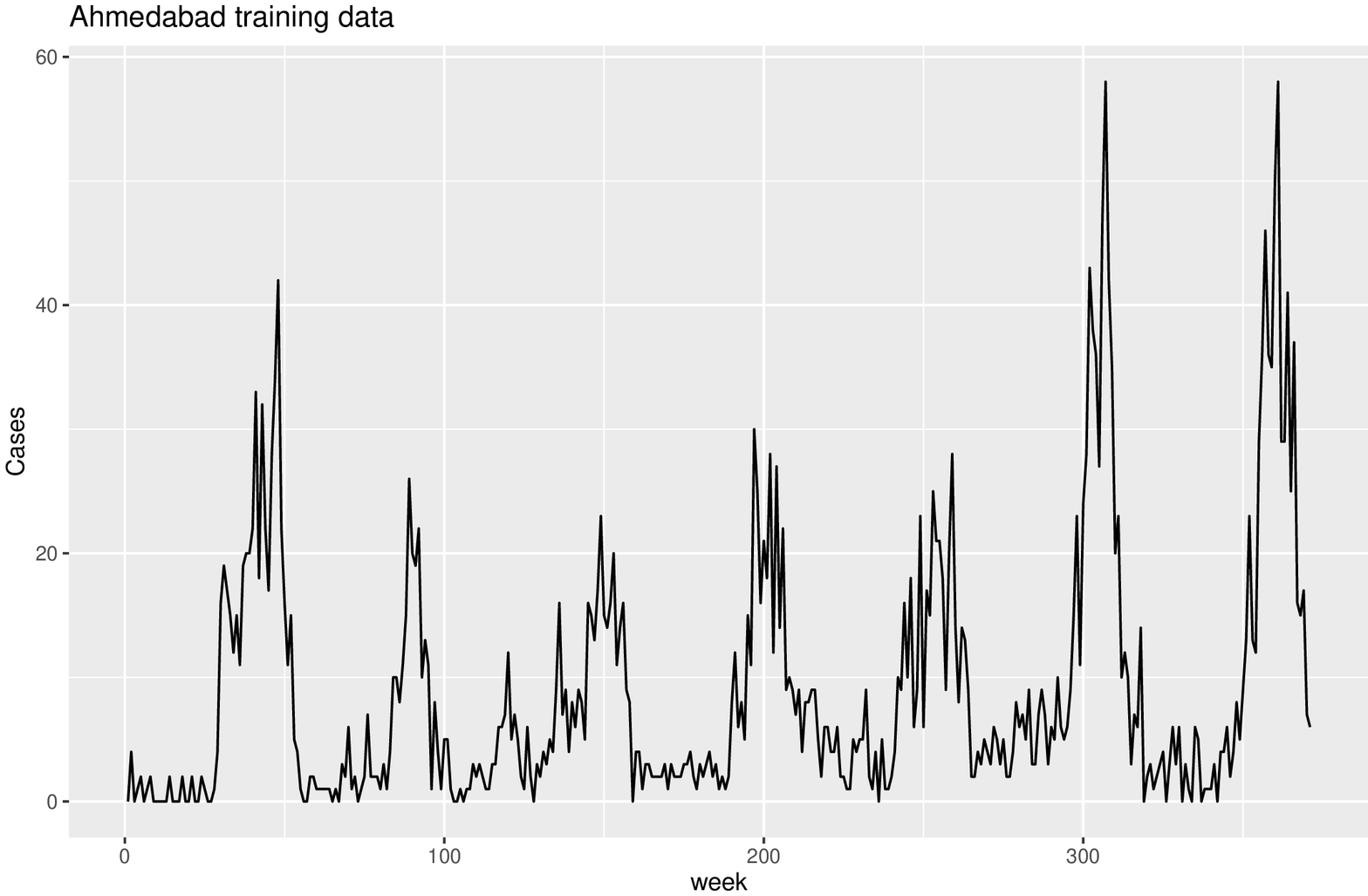}
        \end{minipage}
        &
        \begin{minipage}{.3\textwidth}
            \includegraphics[width=40mm, height=30mm]{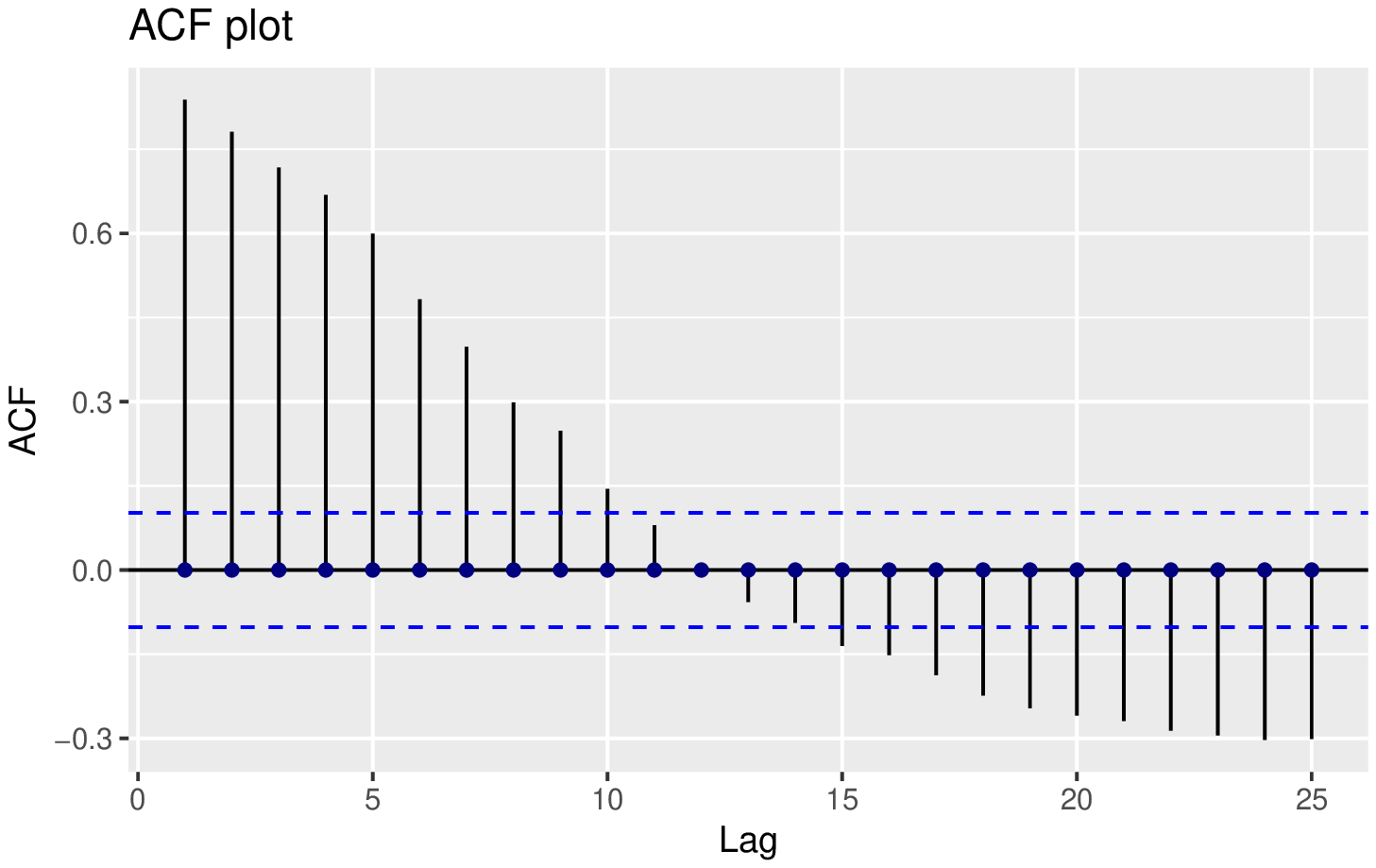}
        \end{minipage}
        &
        \begin{minipage}{.3\textwidth}
            \includegraphics[width=40mm, height=30mm]{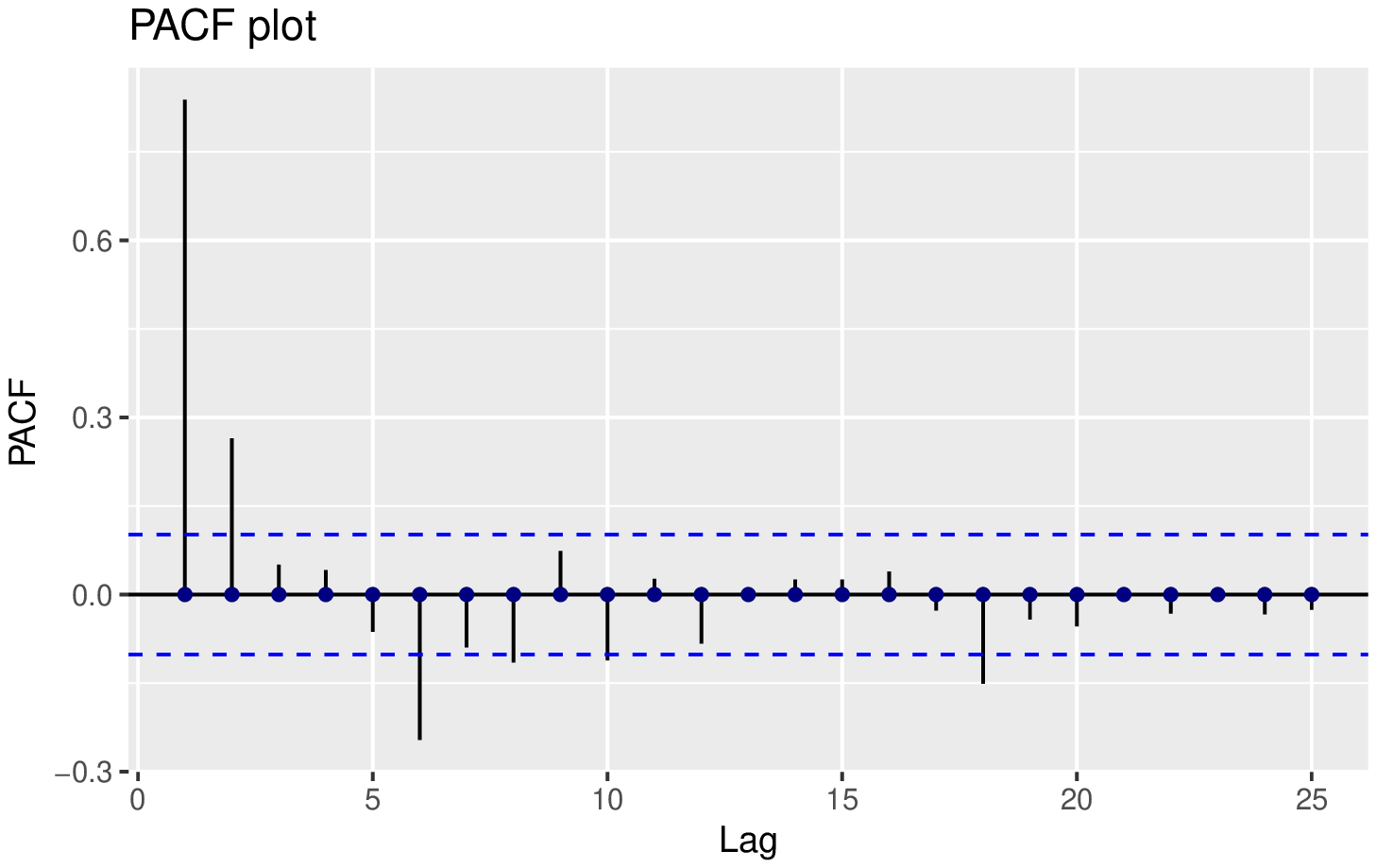}
        \end{minipage}
        \\ \hline
    \end{tabular}
\end{table}

\subsection{Preliminary Data Analysis}
\begin{figure}[ht]
	\centering
	\includegraphics[width=1.0\textwidth]{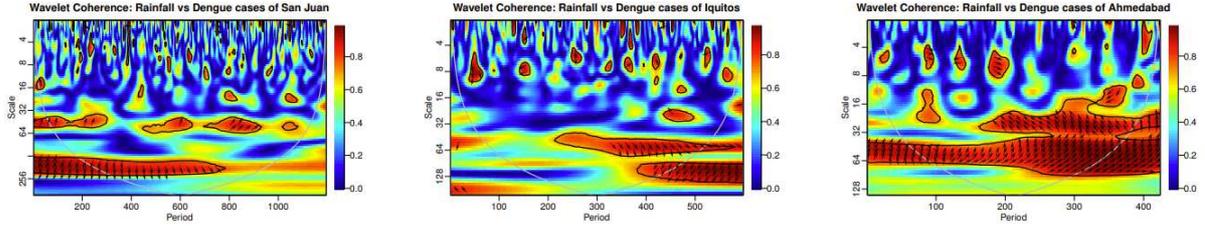}
	\caption{Wavelet coherence plot of San Juan, Iquitos, and Ahmedabad. Colder (blue) to warmer (red) colors indicate an increasingly significant interrelationship between dengue and rainfall. Arrows represent the direction of the relationship, and the cold areas beyond the boundary have no significant relationship.}
	\label{fig:data_wv_coh}
\end{figure}

Dengue surveillance datasets considered in this study are free from any missing observations and are of different lengths. Although it comprises some outliers; however, the values are plausible and thus kept as it is. The autocorrelation function (ACF) and partial autocorrelation function (PACF) plots of the dengue caseloads presented in Table \ref{table_acf_pacf} indicate the presence of serial correlation between the observations and its lagged values. Thus, these plots represent the retrospective nature of dengue cases. Furthermore, we analyze specific statistical features of the dengue datasets to unfold the general structure of the disease outbreak. Based on the recommendations of \cite{de200625}, we examine the stationarity, seasonality, linearity, normality, and long-term dependency of three dengue datasets. We perform Kwiatkowski-Phillips-Schmidt-Shin (KPSS) test to study the stationarity of the dengue cases in the three regions \cite{shin1992kpss}. This non-parametric unit root test checks the stationarity of the observed time series against a deterministic trend. Another essential property of any time process is seasonality, which determines the replicating patterns of a time series within a fixed period. We utilize a combined Ollech and Webel's test to verify the seasonality of the series \cite{ollech2020random}. Subsequently, to examine the linearity of the time series datasets, we used Terasvirta's neural network test \cite{terasvirta1993power}. We also verify whether the distribution of dengue incidence cases follows a normal distribution using the Anderson-Darling test \cite{nelson1998anderson}. Additionally, we check the long-term dependency or self-similarity using the Hurst exponent \cite{hurst1951long}. Furthermore, we employ the non-parametric Granger causality (GC) test \cite{granger1969investigating} to verify the causal relationship between rainfall and dengue outbreak. This statistical hypothesis test evaluates whether rainfall helps to forecast dengue cases by measuring the ability of rainfall data to predict the disease's future dynamics. Alongside the GC test, we also consider the wavelet coherence approach to examine the time-localized interaction between the observed series \cite{gouhier2013package}. The wavelet coherence model decomposes the time-indexed data into time-frequency components and provides useful visualization to identify the oscillatory behavior for the corresponding time series. The results of the above-stated statistical tests presented in Table \ref{table_granger_test} reveal specific exciting properties of the dengue datasets. To begin with, we can infer from the KPSS test results that the dengue series are non-stationary except for Iquitos. Secondly, all the dengue incidence series exhibit non-linear and non-gaussian characteristics as observable from the p-values of the Terasvirta test and Anderson-Darling test, respectively. The results from Ollech and Webel test for seasonality narrates that the dengue incidence of Iquitos exhibits seasonality; however, the data for the other two regions is non-seasonal. We also observe that the Hurst exponent value for all the datasets is $>$ 0.5; hence the dengue incidence cases have similarities to their historical observations. Moreover, for performing the GC test, we use single differencing on the non-stationary incidence datasets to meet the stationarity assumption of the GC test and set the order of the test to two, which suggests up to $2^{nd}$ order of lags to include in the auxiliary variable, i.e., rainfall. The computed p-value from Table \ref{table_granger_test} for the GC test of order two indicates a statistically significant causal relationship between rainfall and dengue cases in San Juan and Iquitos region. However, for the Ahmedabad region, the p-value does not indicate that rainfall Granger cause dengue incidence.  In the case of non-linear and non-stationary data, the GC test sometimes fails to quantify the flow of information between time series. A possible remedy to improve the GC test for non-stationary time series is to increase the differencing order, and lag order in the test \cite{stokes2017study}. In this case, we increase the order of lags to four to verify the causality structure in the oscillatory time series data. From the p-values of the GC test with order four, we can infer a significant causal relationship exists between rainfall and dengue incidence cases for all the datasets. However, a stronger causality check can be produced using a wavelet coherence plot for the non-stationary and non-linear time series of dengue cases vs rainfall \cite{adebayo2022environmental}. 
The wavelet coherence plot in Fig. \ref{fig:data_wv_coh} can be interpreted as follows: ({a}) Time is plotted along the horizontal axis, and the vertical axis represents scale; ({b}) Areas on the time-frequency space where movements of two-time series co-occur are located by the wavelet coherence; ({c}) Significant relationship between the two series are marked with warmer colors (red), whereas, the colder colors (blue) represents low dependency between the series; ({d}) Area on the time-frequency space lying beyond the boundary marked with colder shades signifies no relationship between the observed time series; ({e}) The arrows on the plot indicates the direction of the relationship between the variables such that right-down (right-up) or left-up (left-down) direction symbolizes that the first (second) series is the leading cause for the second (first). Thus, similar to the results of the GC test (with order four), the wavelet coherence plot also depicts that the rainfall and dengue incidence series exhibit similar oscillatory behavior for all the regions. Motivated by these results, we use rainfall data as an exogenous causal variable to forecast dengue cases for these three locations.

\begin{table}
	\centering \caption{Global characteristics of the dengue datasets. In the case of statistical tests, the p-value ($<$ 0.1) within the bracket with $^*$ indicates statistical significance.}\label{table_granger_test}
 \scriptsize
 \makebox[\textwidth]{
	\begin{tabular}{|c|c|c|c|}
		\hline
	   Features and Tests & \textbf{San Juan} & \textbf{Iquitos} & \textbf{Ahmedabad} \\ \hline
	    Length & 1196 (1991-2012) & 598 (2002-2012) & 424 (2005-2012)\\ \hline
	    Train : Test (short-term) & 1170 : 26 & 572 : 26 & 398 : 26\\\hline
	    Train : Test (long-term) & 1144 : 52 & 546 : 52 & 372 : 52 \\\hline
	    KPSS test & Non-stationary ($<$ 0.01$^*$) & Stationary ($>$ 0.1) & Non-stationary (0.02$^*$) \\ \hline
        Terasvirta test & Non-linear (0.032$^*$) & Non-linear ($<$ 0.01$^*$) & Non-linear (0.002$^*$) \\ \hline
        Ollech and Webel test & Non-seasonal & Seasonal & Non-seasonal \\ \hline
        Anderson-Darling test & Non-normal ($<$ 0.01$^*$) & Non-normal ($<$ 0.01$^*$) & Non-normal ($<$ 0.01$^*$) \\ \hline
        Hurst Exponent & Long-term dependent (0.739) & Long-term dependent (0.603) & Long-term dependent (0.673) \\ \hline
        Granger Causality test & Causality found & Causality found  & Causality not found  \\ 
        (order = 2, Cases Vs. Rainfall) & (0.065$^*$) & (0.012$^*$) & (0.573) \\\hline
        Granger Causality test & Causality found & Causality found  & Causality found  \\ 
        (order = 4, Cases Vs. Rainfall) & (0.032$^*$) & (0.030$^*$) & (0.049$^*$) \\\hline
    \end{tabular}
    }
\end{table}

%\begin{figure}[h]
%    \centering
%    \includegraphics[width=0.42\textwidth]{San_Juan_WC_New.eps}
%    \includegraphics[width=0.42\textwidth]{Iqt_WC_Plot.eps}
%    \includegraphics[width=0.42\textwidth]{Ahm_WC_New.eps}
%    \caption{Wavelet coherence plot of San Juan, Iquitos, and Ahmedabad. Colder (blue) to warmer (red) colors indicate an increasingly significant interrelationship between dengue and rainfall. Arrows represent the direction of the relationship, and the cold areas beyond the boundary have no significant relationship.}
%    \label{fig:data_wv_coh}
%\end{figure}

\subsection{Development of Proposed XEWNet Model}
This work introduces an ensemble wavelet neural network with an exogenous factor(s) (XEWNet) framework for forecasting dengue incidence based on past disease dynamics and rainfall patterns. The proposed model overcomes the deficiencies of individual neural networks and improves the forecasting performance by including precipitation data. This section summarizes the notion of the maximal overlap discrete wavelet transformation (MODWT) used in the network, followed by a detailed description of the proposed XEWNet architecture.

\subsubsection{MODWT Algorithm}
Real-world datasets comprise signals and noises whose effective analysis yields valuable information about the generating process. For quantifying these structures present in the signal, a precise mathematical operation is required to look at the data through the noise. Wavelet transform is a powerful signal decomposition technique that uses specialized functions called wavelets to analyze the signal \cite{percival2000wavelet}. The key feature of a wavelet is its short-lived wave-like oscillations which are localized in time. This characteristic enables the wavelet transform to represent a signal in both time and scale (frequency) domains, unlike the Fourier transform. Broadly, there are two types of wavelet transformations: (\textit{i}) Continuous wavelet transform (CWT) and (\textit{ii}) Discrete wavelet transform (DWT). While the former variant allows the wavelet to vary continuously, resulting in an infinite number of wavelets, the latter samples the wavelet discretely w.r.t time and frequency. Hence, our study focuses on the DWT approach since the dengue incidence data is sampled from a finite interval at discrete time points. Use of DWT as a localized approximator can be found in several domains, namely in energy \cite{nazaripouya2016univariate, benaouda2006wavelet}, climatology \cite{nury2017comparative}, material science \cite{li2020comparative}, epidemics \cite{panja2022epicasting}, and economics \cite{ko2015international}. A detailed description of DWT is given in Appendix \ref{DWT_formulation}. Despite having vast applicability, the DWT approach has certain drawbacks, such as restricting the size of the signal to be strictly a power of 2. Moreover, the DWT also downsamples, i.e., a simple shift in the signal can significantly impact the signal energy of the DWT by scale and shift. These hinder the application of DWT in specific real-world situations, and to overcome these drawbacks of the DWT approach, a maximal overlapping version of the DWT (MODWT), also termed as the {non-decimated wavelet transform} has been proposed in the literature \cite{percival2000wavelet, walden2001wavelet, percival1997analysis}. The MODWT algorithm is a linear filtering approach that transforms a signal into a time-dependent wavelet and scaling coefficients related to variations over a set of scales. Unlike the DWT, its maximal overlapping version is well-defined for signals of varying lengths and retains the downsampled values at each stage of decomposition. These appealing properties of MODWT make it ideal for analyzing time series from different domains \cite{li2020comparative, benaouda2006wavelet, wang2002multiple, nury2017comparative}. Our study utilizes the multi-resolution analysis (MRA) of the shift-invariant MODWT approach for transforming non-stationary real-world dengue incidence series. The mathematical formulation of the MODWT algorithm is presented in Appendix \ref{MODWT_MRA_Mathematical}. Since the real-world dengue incidence series exhibits high-frequency components for short time intervals and low-frequency components for a long period, this decomposition technique is more relevant. Thus in our study, the MODWT-based MRA technique has been applied to decompose the non-stationary, long-term dependent series into simpler components and extract the relevant information. The following subsection describes the proposed XEWNet model that integrates the wavelet transformation with an ensemble of neural networks to enhance the dengue outbreak's predictability.

Mathematically, given a dengue incidence time series $Y = \{Y_1,Y_2,\ldots,Y_N\}$ with $N$ historical data points, the MODWT algorithm decomposes the original series into details and smooth coefficients:
\begin{equation}
    Y_t = \sum_{l=1}^L D_{l,t} + S_{L,t},
    \label{MODWT_XEWNet}
\end{equation}
where $D_{l,t}$ denotes details coefficients (irregular fluctuations or high-frequency components) at scale $l$, $(l = 1,2,\ldots,L, 
\; \text{ where } L = \log_e (N)-1)$ and $S_L$ is  the smooth coefficient (overall trend or low-frequency components) of the original series \cite{percival1997analysis}.

\begin{figure}[H]
    \centering
    \includegraphics[width = 0.9\textwidth]{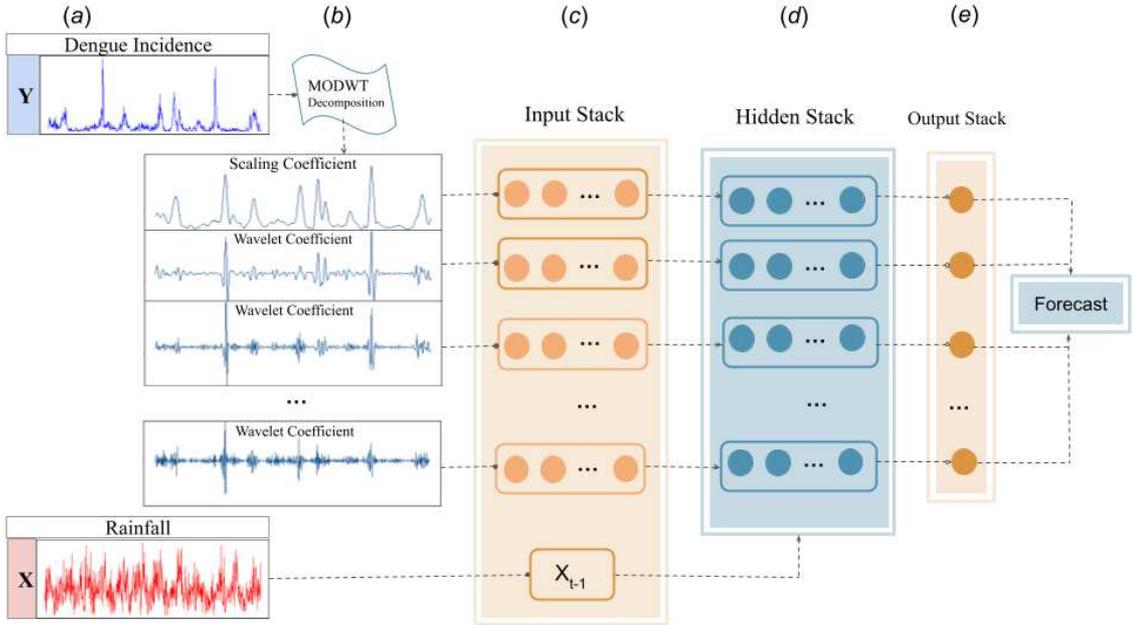}
    \caption{The proposed XEWNet workflow: ({a}) To predict dengue incidence cases, we provide a weekly time series of dengue cases ($Y_t$) and rainfall ($X_t$) in the training period; ({b}) We perform a MODWT based MRA transformation on Y and generate multiple series of details and smooth coefficients; ({c}) We begin to train numerous auto-regressive neural networks to individually model the transformed series along with rainfall dataset in the input stack; ({d}) Each of the neural networks is trained with a single hidden layer having a pre-specified number of nodes inside the hidden stack; ({e}) The output stack comprises of one-step ahead forecast generated by individual neural networks. These predictions are combined to generate the final out-of-sample forecast.}
    \label{XEWnet_Figure}
\end{figure}

\subsubsection{Proposed Model}
In our proposed XEWNet framework, we simultaneously utilize the MODWT algorithm as a pre-processing step to extract the local spectral and temporal information from the input signal. A pivotal characteristic of the multi-resolution analysis of the MODWT approach is that it helps decompose the input signal into non-identical frequency bands exploited for forecasting in the proposed model. 

Real-world dengue incidence series exhibit distinctive characteristics such as non-stationary, non-linearity, long-term dependency, and non-normality in the majority of cases (as reported in Table \ref{table_granger_test}). Thus, to model the underlying mechanism that generates these real-world irregular time series, we transform them using MODWT decomposition with Daubechies orthogonal wavelets of length 2, i.e., the Haar filter. This linear MODWT transformation has time-frequency localization capacity and it can efficiently separate the long-range dependency, non-stationarity, and trends in signals \cite{zhu2014modwt}. Thus the MODWT-based MRA decomposition of the series $Y_t$ generates a set of $L+1$ new uncorrelated random variables, namely the wavelet coefficients. In our proposed methodology, instead of using the actual signal $Y_t$, we model these $L+1$ random variables with distinct neural networks in an ensemble framework. These neural networks' architecture and learning mechanisms enable them to model and control the complexities of the $L+1$ non-linear wavelet coefficients. Therefore, the application of neural networks in the proposed framework can aptly handle the non-linearity present in the three dengue incidence datasets. Moreover, to capture the causal relationship of input series and exogenous rainfall variable ($X_t$), we model $X_t$ with each of the $L+1$ decomposed series of $Y_t$ inside their corresponding network. The forecasts generated from these $L+1$ neural networks are then aggregated to generate the future dynamics of $Y_t$. Thus $h$-step ahead forecasts of $Y_t$ denoted by $\hat{Y}_N^{(h)}$, based on $N$ historical observations, can be mathematically expressed as:
\begin{equation}
    \hat{Y}_N^{(h)}= \sum_{l=1}^L\hat{D}_{l,N}^{(h)} + \hat{S}_{L,N}^{(h)},
    \label{Final_output}
\end{equation}
where $\hat{D}_{l,N}^{(h)} \; (l = 1,2,\ldots, L)$ and $\hat{S}_{L,N}^{(h)}$ are the forecasts generated by $L+1$ neural networks. We call this model combining $L+1$ auto-regressive neural networks in an ensemble setting with an exogenous variable as XEWNet$(p,k)$ model. Thus, our task now reduces to designing an optimal neural network for each transformed series in XEWNet($p,k$) model. In each of the base models in the ensemble, we utilize a feed-forward auto-regressive neural network (ARNN) architecture comprising three layers - one input layer with $(p+1)$ nodes, one hidden layer having $k$ nodes, and an output layer. Each of these neural networks is employed for modeling $p$ lagged observations of the wavelet coefficients and an exogenous variable to generate a one-step forecast of the corresponding series. Hence, the output generated by $L+1$ neural networks with the above-stated architecture can be mathematically expressed as
\begin{align*}
    \hat{D}_{l,t}^{(1)} &=  \beta_{o,l} + \sum_{\tilde{h}=1}^k \gamma_{\tilde{h},l} \phi  \left( \tilde{\beta_{l}} + \sum_{i=0}^{p-1} \alpha_{i,l} D_{l,t-i} + \alpha_{x,l} x_t \right) 
    ; \; (l=1,2,\ldots,L), \\
    \hat{S}_{L,t}^{(1)} &=  \eta_{o} + \sum_{\tilde{h}=1}^k \delta_{\tilde{h},L} \phi  \left( \tilde{\eta_{L}} + \sum_{i=0}^{p-1} \lambda_{i,L} S_{L,t-i} + \lambda_{x,L} x_t \right) ;
\end{align*}
where $\{\beta_{o,l}\}, \{\tilde{\beta}_l \}$ denotes the bias terms, the weights $\{ \alpha_{i,l}, \alpha_{x,l}\}$ are the connections of the lagged inputs and exogenous variable with the hidden neurons, respectively, $\{\gamma^{'}_l \}$ denotes the $k$-dimensional weight vector between the hidden layer and the output layer for the neural networks of the details coefficients. The corresponding parameters of the neural network for modeling the smooth series of $Y_t$ are given by $ \eta_o, \tilde{\eta_L}$ as the bias, $\{\lambda_{i,L}, \lambda_{x,L}\}$ denote the input node to hidden node connections, and $\delta^{'}_L$ denotes the hidden layer to output layer weights. The logistic sigmoidal activation function is denoted by $\phi$ \cite{rumelhart1985learning}. We initialize the connection weights at random starting values and train them using gradient descent back-propagation approach \cite{rumelhart1985learning}. The above-stated procedure generates a one-step-ahead forecast, whereas we feed back the latest forecast and remove the last lagged observation from the model iteratively for computing a multi-step-ahead forecast. Eventually, we forecast the original series by utilizing the outputs from different models, as stated in Eqn. \ref{Final_output}. 

The proposed XEWNet model has three hyperparameters, namely, $L+1$ denoting the number of levels of wavelet decomposition, $p$ indicating the number of lagged observations of the target variable, and $k$ quantifying the total number of hidden neurons in the hidden layer. Previous studies have identified that the essence of the choice of wavelet decomposition level is to reveal the complex variability of the transformed signal \cite{percival1997analysis, aussem1998waveletbased}. MODWT eliminates alignment artifacts, and the decomposition level $L+1$ consists of $(L+1)N$ coefficients rather than the original series of size $N$ (whereas simple DWT has $N$ coefficients for any given $L+1$). The MODWT pyramid algorithm requires $N\log_2(N)$ multiplications, which is the same as needed by the Fast Fourier Transform (FFT) algorithm \cite{percival2000wavelet}. However, during the practical implementation, to reduce the complexity of computing MODWT coefficients and reduction of redundant wavelet coefficients, Percival and Walden \cite{percival2000wavelet} suggested using $N\log_e(N)$ multiplications since the MODWT algorithm requires $\mathcal{O}(N\log N)$ operations similar to FFT algorithm \cite{papadimitriou1979optimality}. Therefore, an increase in the decomposition level may increase the possibility of creating redundant wavelet coefficients and also increase the computational complexity of the model. Also, a reduction in the decomposition levels may reduce the effectiveness of the MODWT approach as opposed to its algorithmic complexity. Hence following this conventional approach, we set $L+1 = \lfloor\log_e\text{(length of training set}) \rfloor$ in our model, where $\lfloor \cdot \rfloor$ denotes the floor function. We utilize the Akaike information criterion (AIC) \cite{deleeuw1992introduction} for determining the best value of $p$, and for stabilizing the neural network we specify $k = \left[\frac{p}{2}+1\right]$ this restricts the over-fitting and under-fitting of the network \cite{panja2022epicasting, hyndman2018forecasting}. A schematic workflow of the proposed XEWNet model is presented in Fig. \ref{XEWnet_Figure}. As the figure shows, the target time series (indicated in blue) is first transformed using a MODWT-based MRA into several levels. The lagged values of these levels and the exogenous variable (marked in red) are passed through the input stack. The connection weights connect each input set to the hidden nodes (dashed line with arrows). The outputs from the hidden stack are processed in the output layer, which eventually generates the one-step ahead forecast from the interconnected network. These predictions from several networks are finally aggregated to generate the XEWNet output. Applying wavelet decomposition to diagnose the main signal from the given series and using an auto-regressive neural network with auxiliary information in an ensemble framework for enhancing the forecast accuracy justifies the nomenclature of the proposed model as XEWNet. The implementation technique of the proposed XEWNet is summarized in Algorithm \ref{alg:cap}.

\begin{algorithm}[H]
%\renewcommand{\thealgorithm}{}
%\begin{algorithmic}
    %\SetKwFunction{isOddNumber}{isOddNumber}
    \SetKwInOut{KwIn}{Input}
    \SetKwInOut{KwOut}{Output}
    \KwIn{ Two in-sample time series ($Y_t,X_t$) of length $N$ denoting the target and exogenous variables.}
    \KwOut{ Forecast of target series for $h$ time steps.}
    %\vspace{0.2cm}
    %\setcounter{AlgoLine}{}
    \begin{enumerate}
        \item Transform the target series, $Y_t$, into corresponding wavelet coefficients by applying multi-resolution analysis of maximal overlap discrete wavelet transformation (MODWT) using a haar filter.
        \item Generate $L+1$ uncorrelated time series objects by extracting $L$ detail series and one smooth series from the above-stated decomposition.
        \item Select $L+1$ as $\log_e (N)$, where $N$ denotes the size of training dataset. 
     \end{enumerate}
    % Transform the target series, $Y_t$, into corresponding wavelet coefficients by applying multi-resolution analysis of maximal overlap discrete wavelet transformation (MODWT) using a haar filter.\\
    % Generate $L+1$ uncorrelated time series objects by extracting $L$ detail series and one smooth series from the above-stated decomposition. \\
    % \nl Select $L+1$ as $\log_e (N)$, where $N$ denotes the size of training dataset. \\
    \For{$l \leftarrow 1$ \KwTo $L$}{
       Model $l^{th}$ detail series with $X_t$ using an auto-regressive neural network. \\
       Design the network with one input layer having $p+1$ nodes, one hidden layer with $k$ neurons, and an output layer. \\
       Input $X_t$ and $p$ lagged values of the $l^{th}$ detail series in the feed-forward architecture, which subsequently passes through $k$ hidden neurons and generates a one-step-ahead forecast.  \\
       Select $p$ by minimizing the Akaike information criterion (AIC) and $k =  \left [\frac{\displaystyle{p}}{\displaystyle{2}}+1 \right ]$, for stable learning architecture. \\
       \KwRet{One-step ahead forecast for the corresponding details series.} \\
    }
    \begin{enumerate}
      \setcounter{enumi}{3}
       \item Model the smooth series of $Y_t$ and $X_t$ with a similar auto-regressive neural network as defined above and generate the subsequent forecast.
        \item Form an ensemble framework to combine the forecasts generated from different series and formulate the final one-step-ahead prediction.
        \item Generate multi-step ahead forecast iteratively by replacing the distant historical observation with the latest forecast.
   \end{enumerate}
    \caption{Proposed Ensemble Wavelet Neural Network with exogenous factor(s) (XEWNet($p,k$))}\label{alg:cap}
    %\end{algorithmic}
\end{algorithm}

\section{Numerical Experiments and Results}\label{Exp_results}
In this section, we verify the practical utility of the proposed XEWNet framework by conducting extensive experimentation on three dengue datasets. In the subsequent subsections, we discuss several accuracy measures used in this study (refer to Section \ref{Perf_Met}), benchmark models from different paradigms (refer to Section \ref{Bselines}), implementation of the proposed architecture (refer to Section \ref{Implement}), and performance evaluation of the proposal in comparison with the state-of-the-art forecasters (refer to Section \ref{Performance}).

\subsection{Performance Metrics} \label{Perf_Met}
This study adopts four popularly used key performance indicators, namely Root Mean Square Error (RMSE), Mean Absolute Error (MAE), Symmetric Mean Absolute Percent Error (SMAPE), and Mean Absolute Scaled Error (MASE), to evaluate the deviation between the forecasts and the ground truth. The formulae for these metrics are as follows:
\begin{align}\label{accuracy_metrics}
    RMSE &= \sqrt{\frac{1}{h} \sum_{i=1}^h (y_i - \hat{y}_i)^2} ; \; \;
    MAE = \frac{\displaystyle 1}{\displaystyle h} \sum_{i=1}^h |y_i - \hat{y}_i|;\\\nonumber
    SMAPE &=\frac{1}{h} \sum_{i=1}^h \frac{|\hat{y}_i - y_i|}{(|\hat{y}_i|+|y_i|)/2} \times 100 ; \; \;
    MASE = \frac{1}{h} \frac{\sum_{i=N+1}^{N+h} |\hat{y}_i - y_i|}{\frac{1}{N-\mathcal{s}}|y_i - y_{i-\mathcal{s}}|};
\end{align}
where $y_i$ and $\hat{y}_i$ are the ground-truth and its corresponding forecast, respectively, $N$ represents the training size, $h$ is the forecast horizon, and $\mathcal{s}$ denotes the seasonality of the series. By general convention, the model having the least values of these metrics is the ``best" performing model \cite{makridakis2020m4}.

\subsection{Benchmark Forecasters} \label{Bselines}
We utilize several state-of-the-art methods from different paradigms to compare the proposed model's effectiveness with other baseline forecasters. This analysis has two primary objectives: firstly, to assess the importance of rainfall data in predicting dengue transmission, and secondly, to check whether an ensemble wavelet framework has the potential to generate more reliable forecasts compared to individual forecasters. To verify these two hypotheses, we compare our proposed XEWNet approach with two types of models. One class of models is trained exclusively with the historical dengue incidence dataset, and the other class of models receives both dengue and rainfall datasets as input; however, their formulation does not possess any hybridization or ensemble framework. In this regard we consider statistical models specifically exponential smoothing (ETS) \cite{ETS}, auto-regressive integrated moving average (ARIMA) \cite{ARIMA}, self-exciting threshold auto-regressive (SETAR) \cite{SETAR}, Theta \cite{THETA}, Wavelet ARIMA (WARIMA) \cite{WARIMA}, TBATS (T: Trigonometric. B: Box-Cox transformation. A: ARIMA errors. T: Trend. S: Seasonal components) \cite{TBATS}, and Bayesian structural time series (BSTS) \cite{BSTS}; machine learning methods, such as artificial neural network (ANN) \cite{ANN} and auto-regressive neural network (ARNN) \cite{faraway1998time}; hybrid models, namely ARIMA-ARNN \cite{chakraborty2021unemployment}, and ARIMA-WARIMA \cite{chakraborty2020real}; and ensemble models viz, ARIMA-ETS-THETA \cite{chakraborty2020nowcasting}, ARNN-THETA-ETS \cite{chakraborty2020nowcasting}, and ensemble wavelet neural network (EWNet) \cite{panja2022epicasting} to forecast the dengue cases based on its historical observations. We further extend the experimentation by providing rainfall data as a covariate to some of the traditional models like exponential smoothing with exogenous variables (ETSX) \cite{hyndman2018forecasting} and auto-regressive integrated moving average with exogenous variables (ARIMAX) \cite{hyndman2018forecasting}; auto-regressive neural network with exogenous variables (ARNNX) \cite{hyndman2018forecasting}; and deep learning frameworks including Neural basis expansion analysis with exogenous variables (NBeatsX) \cite{NBEATS}, Temporal convolutional networks with exogenous variables (TCNX) \cite{TCN}, Transformers with exogenous variables (TransformersX) \cite{TRANSFORMERS}, and Block recurrent neural network with exogenous variables (BlockRNNX) \cite{herzen2022darts}. We summarize the implementation technique for the above-stated models in the subsequent section.

\subsection{Model Implementation} \label{Implement}
The implementation of the proposed XEWNet framework is done using R statistical software. To begin with, we partition the dengue and rainfall datasets into training (in-sample) and test (out-of-sample) segments. Two test sets of fixed lengths (26 weeks and 52 weeks) are considered in this study to extrapolate the forecast efficiency of the proposal for different horizons. Given the in-sample data we employ MODWT based MRA approach with `haar' filter on the dengue dataset using \textit{modwt} function of ``wavelets" package in R. We specify the number of levels of decomposition as $L+1 = \left [ \log_e (\text{ length of training set  }) \right ]$ and boundary condition as periodic. This transformation generates $L$ wavelet coefficients (details) series and one scale coefficients (smooth) series of the original dengue dataset. In the followed-up step, we deploy several auto-regressive neural networks locally to model the transformed series and rainfall data as an auxiliary variable using the \textit{nnetar} function of the ``forecast" package in R. To restrict the redundancy of neural networks; we specify the architecture of the network as having $p+1$ neurons in the input layer and $k$ hidden nodes distributed in a single hidden layer. The size of the input layer ($p+1$) is determined using the minimum AIC and $k$ is specified as $\left [ \frac{\displaystyle{p}}{\displaystyle{2}} + 1 \right ]$. Thus, each of these $L+1$ neural nets models $p$ lagged values of the transformed series and rainfall as a covariate to generate a one-step forecast. Eventually, we combine the local predictions from different models and generate the final (ensemble) output utilizing an ensemble approach. This method is then used iteratively to generate a multi-step ahead forecast. These out-of-sample forecasts of the desired length are subsequently compared with the ground truth using the accuracy metrics defined in Eqn. \ref{accuracy_metrics} and the corresponding values pertaining to specific regions are reported in Tables \ref{sanjuan_accuracy_table_1_year} - \ref{ahmedabad_accuracy_table_1_year}. For the implementation of the baseline forecasters like ETS, ETSX, ARIMA, ARIMAX, TBATS, Theta, ARNN, and ARNNX, we have used the ``forecast" package of R software. Other models such as SETAR, WARIMA, and ANN are trained using the ``tsdyn", ``WaveletARIMA", and ``nnfor" packages in R. In case of hybrid ARIMA-ARNN and ARIMA-WARIMA we have utilized the implementation provided in \cite{chakraborty2019forecasting} and \cite{chakraborty2020real}, respectively. The deep learning prototypes were executed using the ``darts" library in Python \cite{herzen2022darts}. Lastly, ensemble methods are trained using the standard implementation details available in \cite{chakraborty2020nowcasting, panja2022epicasting}.

\begin{table}
	\centering \caption{Performance metrics with 1 year and 6 months test data set for San Juan. Parameters of XEWNet ($p,k$) are reported for 52 weeks-26 weeks.}\label{sanjuan_accuracy_table_1_year} 
 \scriptsize
 \makebox[\textwidth]{
	\begin{tabular}{|c|c|c|c|c|c|c|c|c|c|c|c|c|c|c|}
		\hline
		\multirow{2}{*}{Paradigm} & \multirow{2}{*}{Model} & \multicolumn{4}{c|}{52 weeks forecast} & \multicolumn{4}{c|}{26 weeks forecast}
        \\ \cline{3-6} \cline{7-10} & & RMSE & MAE & SMAPE & MASE & RMSE & MAE  & SMAPE & MASE  \\ \hline
		\multirow{9}{*}{Statistical} 
		              & ETS \cite{ETS} & 86.28 & 57.17  & 85.78 & 04.71 & 12.78 & 11.48 & 54.64 & 01.95 \\
		\multirow{9}{*}{ Models}    
		              & ETSX \cite{hyndman2018forecasting} & 87.17 & 57.53 & 86.17 & 04.74 & 12.14 & 10.63 & 51.64 & 01.80 \\
				      & ARIMA \cite{ARIMA} & 80.18 & 53.94 & 80.43 & 04.45 & 17.29 & 16.01 & 67.35 & 02.72 \\
				      & ARIMAX \cite{hyndman2018forecasting} & 80.09 & 53.90  & 80.36 & 04.44 & 17.25 & 15.96 & 67.19 & 02.71 \\ 
				      & SETAR \cite{SETAR} & 85.37 & 52.23  & 83.31 & 04.64 & 12.14 & 10.80 & 52.21 & 01.83 \\
				      & Theta \cite{THETA}  & 87.13 & 57.58  & 86.39 & 04.75 & 12.32 & 10.88 & 52.42 & 01.84 \\
				      & WARIMA \cite{WARIMA} & 84.46 & 55.77 & 83.10 & 04.60 & 19.06 & 17.52 & 70.80 & 02.98  \\
				      & TBATS \cite{TBATS}  & 86.27 & 57.17  & 85.78 & 04.71 & 12.79 & 11.49 & 54.65 & 01.95 \\
				      & BSTS \cite{BSTS}  & 82.26 & 54.33 & 80.38 & 04.48 & 14.73 & 13.52 & 60.56 & 02.30 \\ \hline
		\multirow{2}{*}{Machine Learning}
		              & ANN \cite{ANN}  & 74.20 & 45.59 & 72.73 & 04.09 & 17.86 & 16.43 & 68.14 & 02.79\\
		\multirow{2}{*}{Models}
		              & ARNN \cite{ARNN} & 86.65 & 57.76 & 86.21 & 04.70 & 12.42 & 08.26 & 42.90 & 01.40 \\
				      & ARNNX  & 87.59 & 59.40  & 97.85 & 04.90 & 14.74 & 09.55 & 53.11 & 01.62\\ \hline
		\multirow{3}{*}{Deep Learning}
		              & NBeatsX \cite{NBEATS} & 104.6 &	82.58 &	117.9 & 05.14 & 81.54 & 69.33 & 65.19 & 03.52\\
		\multirow{3}{*}{Models}
		              & TCNX \cite{TCN} & 119.7 & 97.72 & 172.4 & 06.09 & 137.4 & 125.4 & 157.1 & 06.37\\
		              & TransformersX \cite{TRANSFORMERS} & 73.99 & 49.61 & 52.68 & 03.47 & 90.06 & 74.05 & 60.91 & 03.76\\
		              & BlockRNNX \cite{herzen2022darts} & 78.27 & 55.39 & 57.99 & 03.47 & 111.3 & 93.92 & 84.14 & 04.77\\ \hline
		\multirow{1}{*}{Hybrid}           
		              & ARIMA-ARNN \cite{chakraborty2019forecasting} & 80.04 & 53.91 & 80.41 & 04.43 & 17.49 & 16.18 & 67.77 & 02.75 \\ 
	    \multirow{1}{*}{Models}
	                  & ARIMA-WARIMA \cite{chakraborty2020real} & 80.39 & 54.10 & 80.58 & 04.46 & 17.45 & 16.02 & 67.07 & 02.75 \\ \hline
	    \multirow{2}{*}{Ensemble}               
	                  & ARIMA-ETS-Theta \cite{chakraborty2020nowcasting} & 84.43 & 56.05 & 83.65 & 04.62 & 13.67 & 12.59 & 58.18 & 02.14 \\ 
	    \multirow{2}{*}{Models}
	                  & ARNN-Theta-ETS \cite{chakraborty2020nowcasting} & 88.44 & 58.38 & 87.98 & 04.81 & 12.02 & 10.35 & 50.63 & 01.76 \\ 
		              & EWNet \cite{panja2022epicasting} & 71.52 & 45.04  & 58.70 & 03.71 & 08.62 & 06.94 & 36.06 & 01.18\\ \hline
		\multirow{1}{*}{Proposed Model}              
		              & XEWNet (10,6)-(10,6) & \textbf{68.49} & \textbf{42.14}  & \textbf{51.67} & \textbf{03.47} & \textbf{07.69} & \textbf{05.66} & \textbf{29.08} & \textbf{0.960}\\  \hline
	\end{tabular}}
\end{table}

\begin{table}
	\centering \caption{Performance metrics with 1 year and 6 months test data set for Iquitos. Parameters of XEWNet ($p,k$) are reported for 52 weeks - 26 weeks.}\label{iquitos_accuracy_table_1_year}
 \scriptsize
 \makebox[\textwidth]{
	\begin{tabular}{|c|c|c|c|c|c|c|c|c|c|c|c|c|c|c|}
		\hline
		\multirow{2}{*}{Paradigm} & \multirow{2}{*}{Model} & \multicolumn{4}{c|}{52 weeks forecast} & \multicolumn{4}{c|}{26 weeks forecast}
        \\ \cline{3-6} \cline{7-10} & & RMSE & MAE & SMAPE & MASE & RMSE & MAE  & SMAPE & MASE  \\ \hline
		\multirow{9}{*}{Statistical}
	                    & ETS \cite{ETS} & 04.86 & 02.68  & 97.25 & 01.32 & 01.86 & 01.64 & 85.85 & 01.14 \\ 
	    \multirow{9}{*}{ Models}
	                    & ETSX \cite{hyndman2018forecasting} & 04.86 & 02.68 & 97.23 & 01.32 & 01.86 & 01.64 & 85.86 & 01.14\\ 
	                    & ARIMA \cite{ARIMA} & 07.19 & 06.52 & 126.4 & 03.23 & 06.34 & 05.98 & 126.8 & 04.15\\
	                    & ARIMAX \cite{hyndman2018forecasting} & 07.17 & 06.50 & 126.5 & 03.22 & 06.34 & 05.97 & 127.0 & 04.14\\
	                    & SETAR \cite{SETAR} & 05.82 & 04.66 & 112.4 & 02.31 & 04.31 & 03.74 & 107.1 & 02.60 \\
	                    & Theta \cite{THETA} & 04.87 & 02.67 & 97.35 & 01.32 & 01.85 & 01.63 & 85.81 & 01.13 \\
	                    & WARIMA \cite{WARIMA} & 08.69 & 07.71 & 126.5 & 03.82 & 07.20 & 06.07 & 118.8 & 04.21 \\
	                    & TBATS \cite{TBATS} & 07.76 & 07.19 & 129.7 & 03.56 & 06.97 & 06.56 & 129.7 & 04.55 \\
	                    & BSTS \cite{BSTS} & 08.06 & 06.37 & 122.7 & 03.15 & 02.37 & 02.08 & 91.19 & 01.44 \\ \hline
	  \multirow{2}{*}{Machine Learning}
	                    & ANN \cite{ANN} & 05.58 & 04.33  & 109.8 & 02.22 & 05.15 & 04.42 & 110.8 & 03.06\\
	  \multirow{2}{*}{Models} 
	                    & ARNN \cite{ARNN} & 05.08 & 02.52 & 103.7 & 01.25 & 01.63 & 01.36 & 86.52 & 0.940\\
	                    & ARNNX \cite{hyndman2018forecasting} & 12.21 & 08.32 & 121.8 & 01.91 & \textbf{01.62} & \textbf{01.36} & 86.52 & \textbf{0.940} \\ \hline
	\multirow{3}{*}{Deep Learning}
		              & NBeatsX \cite{NBEATS} & 12.62 & 8.44 & 110.2 & 01.93 & 13.94 & 11.01 & 114.4 & 01.77\\
	\multirow{3}{*}{Models}
		              & TCNX \cite{TCN} & 44.48 & 29.93 & 145.5 & 06.87 & 32.39 & 22.11 & 91.99 & 03.56 \\
		              & TransformersX \cite{TRANSFORMERS} & 11.95 & 08.05 & 119.9 & 01.85 & 08.52 & 06.71 & \textbf{62.68} & 01.08 \\
		              & BlockRNNX \cite{herzen2022darts} & 12.55 & 08.82 & 129.4 & 02.02 & 13.08 & 10.34 & 86.81 & 01.66 \\ \hline
	\multirow{1}{*}{Hybrid}           
		              & ARIMA-ARNN \cite{chakraborty2019forecasting} & 06.70 & 05.89 & 122.3 & 02.91 & 05.70 & 05.29 & 121.7 & 03.67 \\ 
	\multirow{1}{*}{Models}
	                  & ARIMA-WARIMA \cite{chakraborty2020real} & 06.94 & 06.17 & 123.3 & 03.05 & 05.82 & 05.29 & 120.3 & 03.67 \\ \hline
	\multirow{2}{*}{Ensemble}               
	                  & ARIMA-ETS-Theta \cite{chakraborty2020nowcasting} & 05.11 & 03.54 & 104.3 & 01.75 & 03.09 & 02.67 & 97.75 & 01.85 \\ 
	\multirow{2}{*}{Models}
	                  & ARNN-Theta-ETS \cite{chakraborty2020nowcasting} & 04.87 & 02.90 & 99.07 & 01.43 & 02.21 & 01.94 & 89.54 & 01.35 \\ 
		              %& ARFIMA-Theta-ETS & 5.12 & 3.55 & 104.40 & 1.76 & 3.11 & 2.68 & 97.88  & 1.86 \\ 
		              & EWNet \cite{panja2022epicasting} & 04.78 & 03.01 & 99.30 & 01.49 & 02.25 & 01.95 & 89.10 & 01.35 \\ \hline
	\multirow{1}{*}{Proposed Model}              
		              & XEWNet (7,5)-(1,2) & \textbf{04.73} & \textbf{02.50}  & \textbf{96.41} & \textbf{01.25} & 01.98 & 01.57 & 79.95 & 01.09\\ \hline
	\end{tabular}}
\end{table}

\begin{table} 
	\centering \caption{Performance metrics with 1 year and 6 months test data set for Ahmedabad. Parameters of XEWNet ($p,k$) are reported for 52 weeks-26 weeks.}\label{ahmedabad_accuracy_table_1_year}
 \scriptsize
 \makebox[\textwidth]{
	\begin{tabular}{|c|c|c|c|c|c|c|c|c|c|c|c|c|c|c|}
		\hline
		\multirow{2}{*}{Paradigm} & \multirow{2}{*}{Model} & \multicolumn{4}{c|}{52 weeks forecast} & \multicolumn{4}{c|}{26 weeks forecast}
        \\ \cline{3-6} \cline{7-10} & & RMSE & MAE & SMAPE & MASE & RMSE & MAE  & SMAPE & MASE  \\ \hline
		\multirow{9}{*}{Statistical}
	                    & ETS  \cite{ETS} & 14.26 & 10.18  & 101.4 & 01.98 & 05.48 & 05.22 & 113.3 & 03.96 \\ 
	    \multirow{9}{*}{ Models}
	                    & ETSX \cite{hyndman2018forecasting}  & 14.26 & 10.18  & 101.5 & 01.98 & 05.48 & 05.22 & 113.3 & 03.95\\ 
	                    & ARIMA \cite{ARIMA} & 15.01 & 09.91 & 99.18 & 01.93 & 03.54 & 03.26 & 94.05 & 02.47\\
	                    & ARIMAX \cite{hyndman2018forecasting} & 15.03 & 09.91 & 99.14 & 01.93 & 03.52 & 03.23 & 93.71 & 02.45 \\
	                    & SETAR \cite{SETAR} & 11.79 & 09.64 & 94.55 & 01.88 & 08.18 & 07.62 & 125.9 & 05.77 \\
	                    & Theta \cite{THETA} & 14.05 & 10.09  & 100.4 & 01.97 & 05.66 & 05.40 & 114.6 & 04.09 \\
	                    & WARIMA \cite{WARIMA} & 10.51 & 09.36 & 99.32 & 01.83 & 08.95 & 08.20 & 129.1 & 06.21 \\
	                    & TBATS \cite{TBATS} & 14.26 & 10.18 & 101.5 & 01.99 & 05.48 & 05.22 & 113.2 & 03.95 \\
	                    & BSTS \cite{BSTS}  & 24.90 & 17.21  & 165.9 & 03.36 & 67.06 & 57.98 & 198.4 & 43.92 \\ \hline
	  \multirow{2}{*}{Machine Learning}
	                    & ANN \cite{ANN} & 14.95 & 10.02  & 100.5 & 01.84 & 08.70 & 08.15 & 129.1 & 06.17 \\
	  \multirow{2}{*}{Models} 
	                    & ARNN \cite{ARNN} & 17.86 & 11.40 & 156.8 & 02.22 & 02.05 & 02.69 & 119.5 & 01.28 \\
	                    & ARNNX \cite{hyndman2018forecasting} & 17.29 & 10.96  & 143.5 & 02.14 & 02.07 & 02.71 & 122.7 & 01.29 \\ \hline
	\multirow{3}{*}{Deep Learning}
		              & NBeatsX \cite{NBEATS} & 10.64 & 06.72 & 92.91 & 01.92 & 09.32 & 06.84 & \textbf{38.44} & \textbf{0.760} \\
	\multirow{3}{*}{Models}
		              & TCNX \cite{TCN} & 11.37 & 07.02 & 78.70 & 01.37 & 14.56 & 11.87 & 64.47 & 01.31 \\
		              & TransformersX \cite{TRANSFORMERS} & 09.99 & 06.91 & 78.99 & 01.35 & 11.91 & 08.91 & 48.24 & 0.990 \\
		              & BlockRNNX \cite{herzen2022darts} & 11.37 & 07.03 & 78.70 & 01.37 & 14.56 & 11.87 & 64.47 & 01.31 \\ \hline
	\multirow{1}{*}{Hybrid}           
		              & ARIMA-ARNN \cite{chakraborty2019forecasting} & 15.00 & 09.91 & 99.20 & 01.93 & 03.55 & 03.27 & 94.22 & 02.48 \\ 
	\multirow{1}{*}{Models}
	                  & ARIMA-WARIMA \cite{chakraborty2020real} & 15.21 & 09.91 & 100.8 & 01.93 & 03.60 & 03.20 & 97.88 & 02.47 \\ \hline
	\multirow{2}{*}{Ensemble}               
	                  & ARIMA-ETS-Theta \cite{chakraborty2020nowcasting} & 14.40 & 10.06 & 100.8 & 01.96 & 04.88 & 04.63 & 108.4 & 03.50 \\ 
	\multirow{2}{*}{Models}
	                  & ARNN-Theta-ETS \cite{chakraborty2020nowcasting} & 14.62 & 10.03 & 100.8 & 01.95 & 04.30 & 04.07 & 103.2 & 03.08 \\ 
		              %& ARFIMA-Theta-ETS & 13.96 & 10.10 & 100.43 & 1.97 & 5.38 & 5.57 & 115.85 & 4.22 \\ 
		              & EWNet \cite{panja2022epicasting} & 10.45 & 06.71  & 86.83 & 01.31 & 11.74 & 09.25 & 54.16 & 01.02 \\ \hline
	\multirow{1}{*}{Proposed Model}              
		              & XEWNet (10,6)-(16,9) & \textbf{09.98} & \textbf{06.55}  & \textbf{78.34} & \textbf{01.28} & \textbf{02.04} & \textbf{02.36} & 128.5 & 01.79\\ \hline
	%	ARFIMA(1,0.096,1)  & 13.61 & 10.09  & 99.90 & 1.97 & 6.37 & 6.08 & 119.21 & 4.61 \\ \hline
	\end{tabular}}
\end{table}

\subsection{Model Performance Comparisons} \label{Performance}
This section elaborates on the efficiency of the proposed XEWNet model and other benchmark forecasters to serve as an early warning system for dengue outbreaks. The experimental results obtained by implementing the above-stated models for San Juan, Iquitos, and Ahmedabad regions are reported in Tables \ref{sanjuan_accuracy_table_1_year} - \ref{ahmedabad_accuracy_table_1_year}, respectively. From the accuracy measures for the San Juan region (Table \ref{sanjuan_accuracy_table_1_year}), it is evident that our proposed architecture outperforms all other baseline models for both the forecast horizons (short and long-term trajectories). Particularly, for short-term prediction, the XEWNet framework is 18\% more accurate than the second-best performing forecaster, EWNet. This improvement in forecast accuracy is predominantly due to the use of rainfall data as a covariate. We observe this phenomenon for ARNNX, which generates the most reliable dengue forecast among the baselines with covariates, but its performance drops significantly w.r.t. our proposed model. The performance of the deep learning frameworks is not satisfactory for 26 weeks of prediction; their long-term forecasts are competitive with the proposed model. 

\begin{figure}[H]
    \centering
    \includegraphics[width=0.32\textwidth]{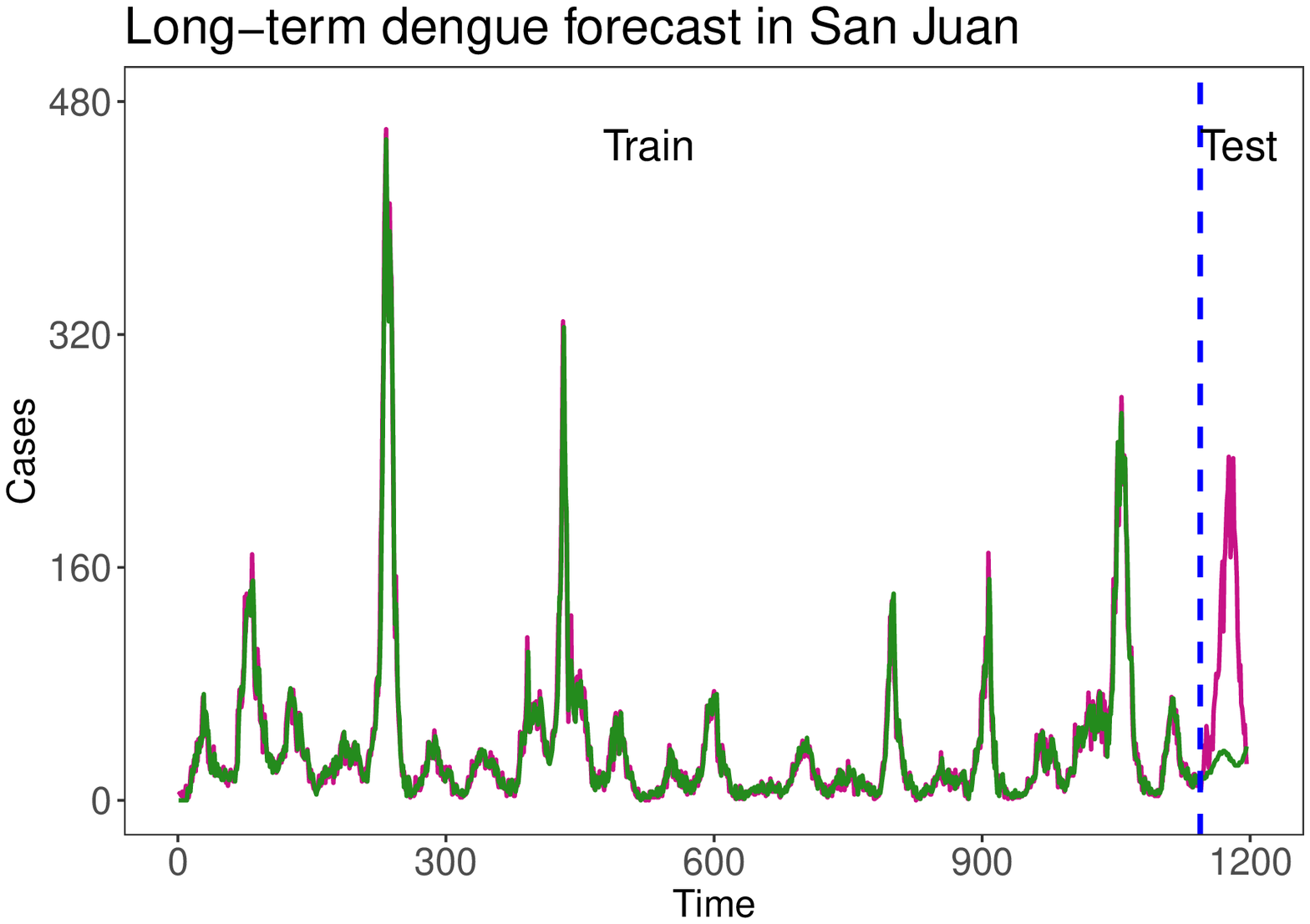}
    \includegraphics[width=0.32\textwidth]{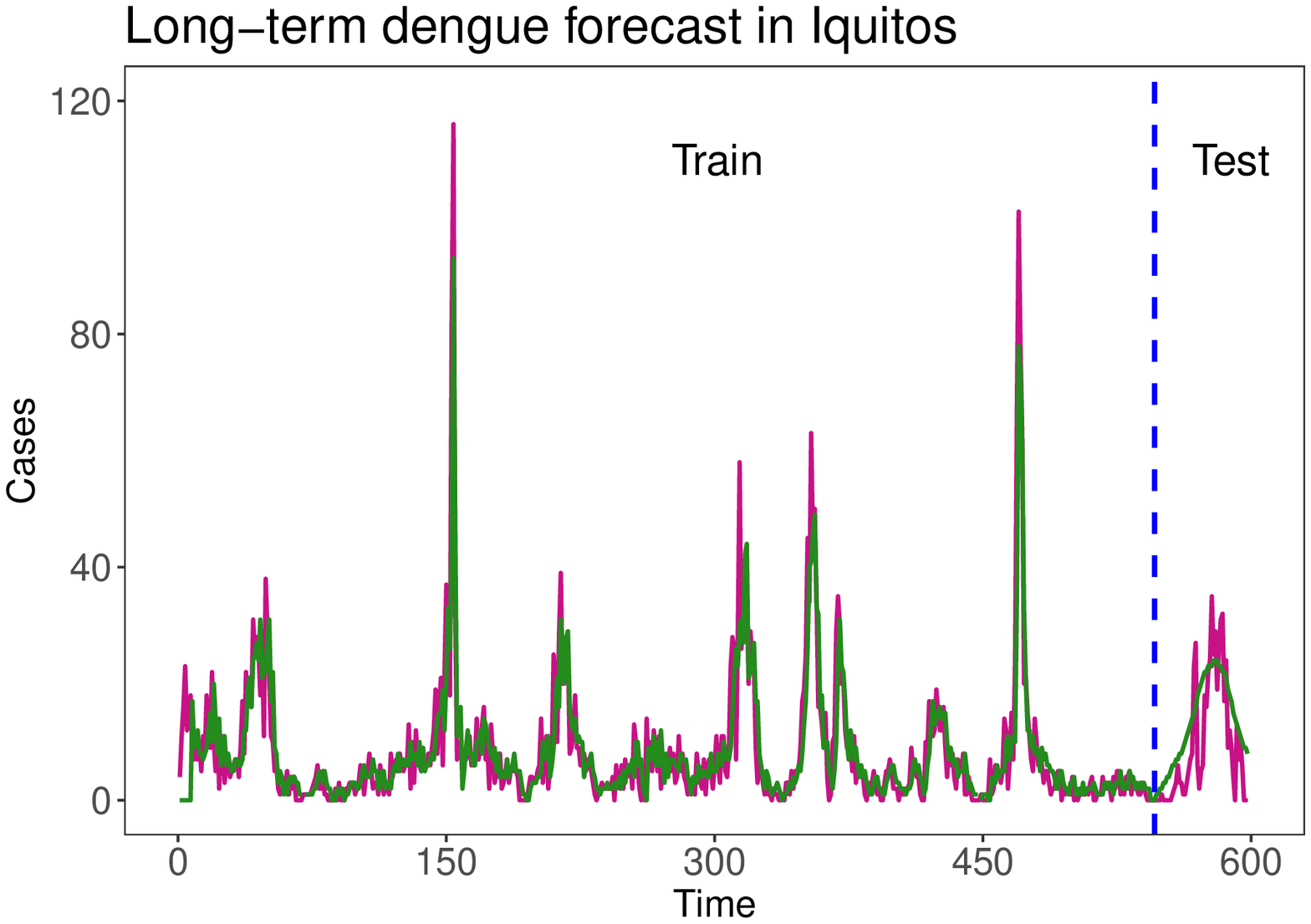}
    \includegraphics[width=0.32\textwidth]{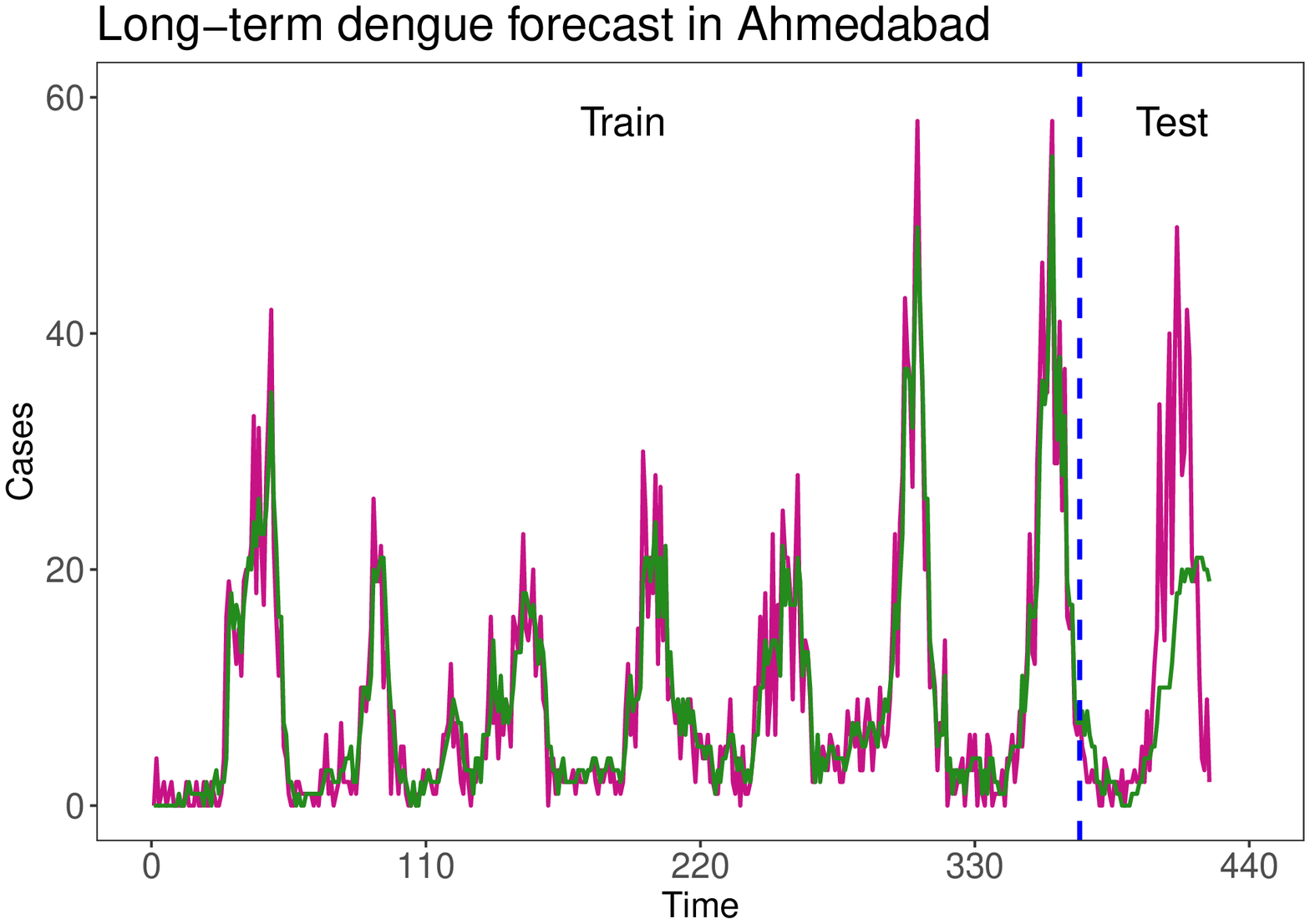}
    \caption{Long-term (52 weeks) dengue cases forecast (green) generated by proposed XEWNet and ground truth data (pink) for the three geographical regions in the study. The blue dashed line marks the distinction between the training and test set.}
    \label{fig:Forecast}
\end{figure}

\begin{figure}[H]
    \centering
    \includegraphics[width=0.32\textwidth]{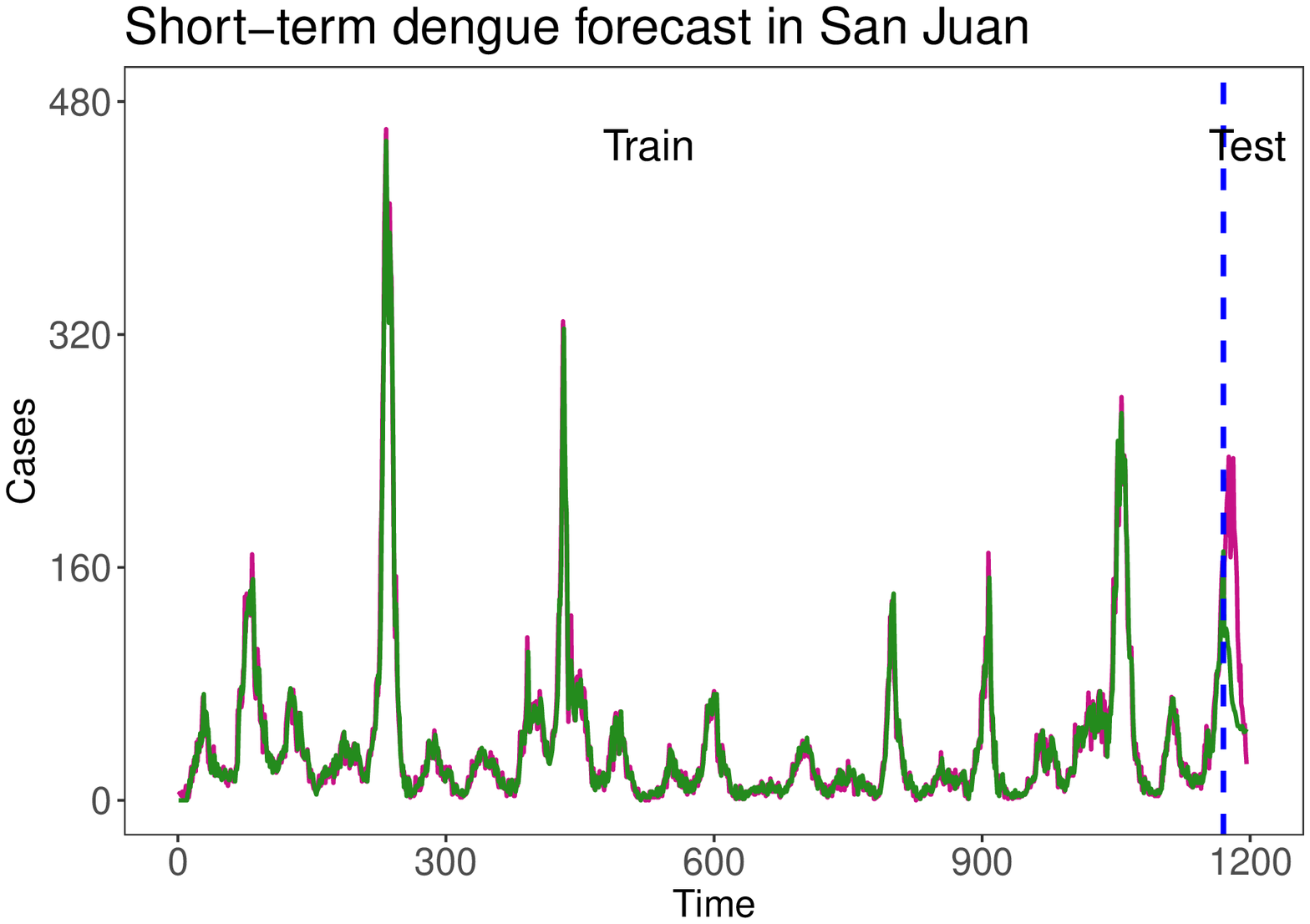}
    \includegraphics[width=0.32\textwidth]{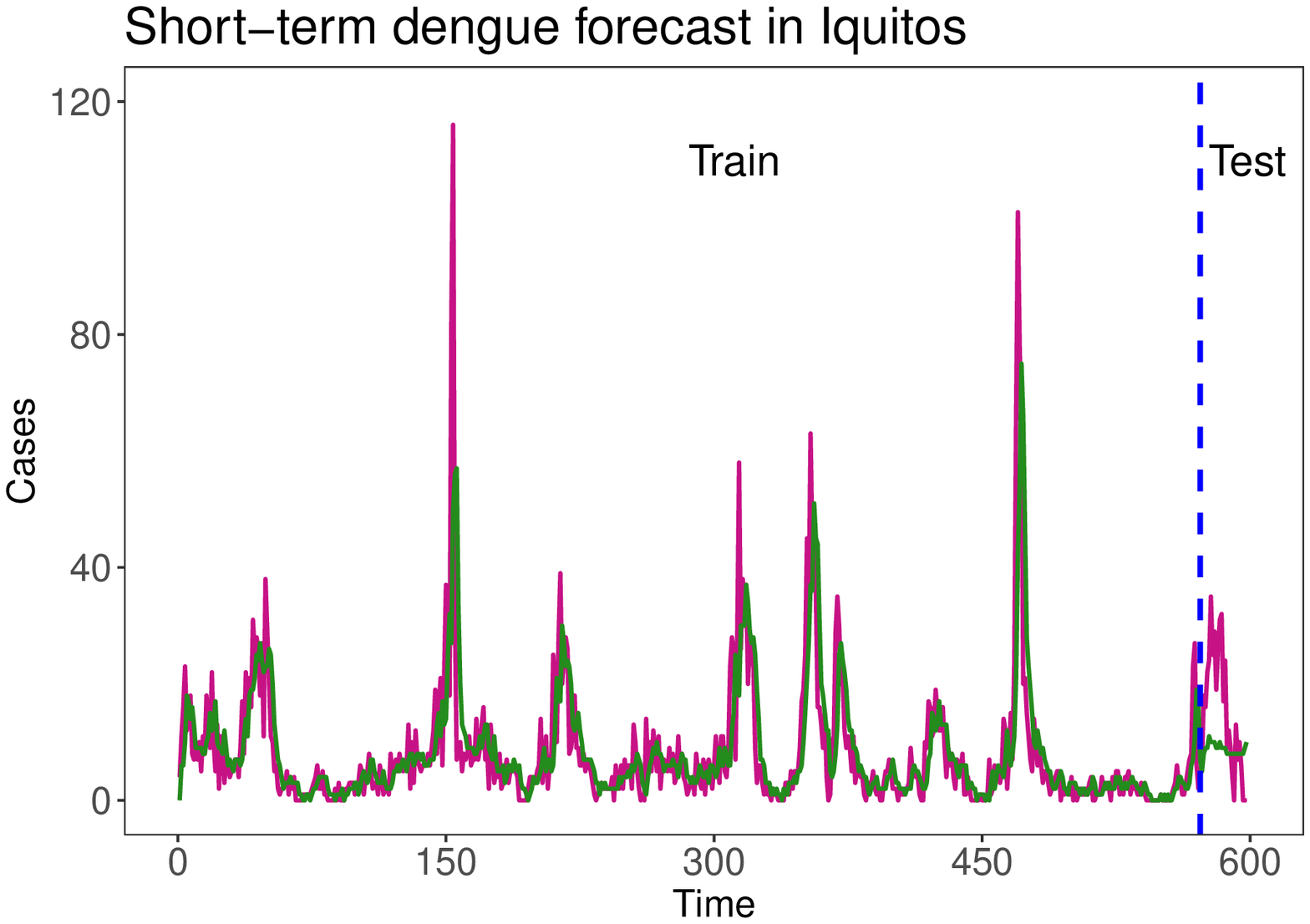}
    \includegraphics[width=0.32\textwidth]{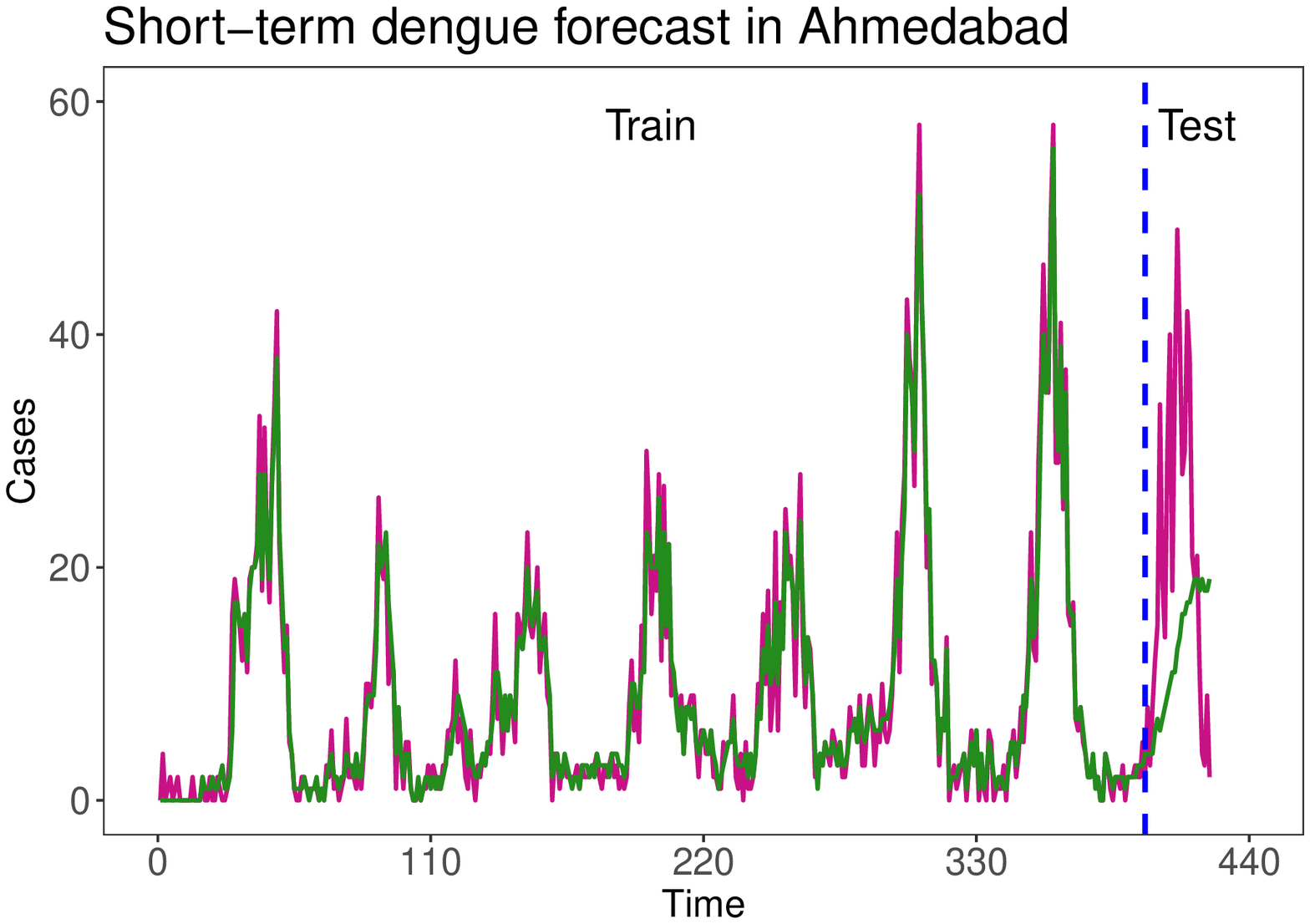}
    \caption{Short-term (26 weeks) dengue cases forecast (green) generated by proposed XEWNet and ground truth data (pink) for the three geographical regions in the study. The blue dashed line marks the distinction between the training and test set.}
  \label{Short-term-diagram}
\end{figure}

In the case of dengue incidence of Iquitos, the short-term predictions generated by the attention-mechanism-based TransformersX and ARNNX frameworks surpass all other baselines. However, these models generate less accurate outcomes for long-term forecasting compared to the proposed XEWNet approach which generates consistently better long-term out-of-sample predictions. The efficiency of the wavelet transformation is found in extracting signal from noise, resulting in reliable long-term predictions for Iquitos data. For the dengue forecasting task of the Ahmedabad region, we observe that the deep learning-based models provide competitive forecasts with 26 weeks lead time. The short-term dengue predictions for this region produced by the XEWNet framework transcends all the models from different paradigms w.r.t the RMSE and MAE scores. However, the other two accuracy measures of the NBeatsX model are minimum. Based on the accuracy metrics for long-term forecasts of this region (Table \ref{ahmedabad_accuracy_table_1_year}), we can detect that TransformersX and EWNet perform significantly better than the benchmark methods. However, compared to the proposed XEWNet framework, their forecasting ability diminishes by 9\%. We diagrammatically present the long-term forecasts generated by the proposed XEWNet model for the three geographical regions in Fig. \ref{fig:Forecast}. The short-term (26 weeks) prediction of the proposed forecaster is provided in Fig. \ref{Short-term-diagram}. Although the experimental results suggest that the proposed XEWNet method generates more accurate dengue forecast for the three regions in comparison to the baseline forecasters as expressed by the evaluation metrics, the plots in Fig. \ref{fig:Forecast} and \ref{Short-term-diagram} demonstrates that in some instances the out-of-sample point estimates generated by the XEWNet model cannot be well fitted with real incidence cases.  
However, our task is to predict the number of dengue cases each week (in each location) based on the climatic condition and an understanding of the relationship between climate and dengue dynamics has been established in this study. This can improve research initiatives and resource allocation to help fight life-threatening epidemics with our proposed early warning system.

The overall experimental evaluation of the proposed model and the benchmark forecasters reveal some interesting observations. The accuracy metrics for linear statistical models like ETS and ARIMA suggest their incapability to handle the irregularities of real-world dengue datasets. Moreover, the inclusion of rainfall data as an exogenous variable marginally improves the forecast accuracy of these traditional models. This failure of ETSX and ARIMAX is primarily attributed to approximating the complex relationship between rainfall and dengue incidence cases by a linear function with a constant rate of change. Although these problems are addressed more precisely by the deep learning frameworks, their overall performance is unsatisfactory for long-term forecasts. The limited availability of historical data is one of the potential reasons for the collapse of these advanced models. Data set size creates a barrier to the performance of the deep learners; however, our proposed XEWNet approach can optimally model the complex non-linear relationship between the observed and covariate series, thus resulting in improved predictions. Unlike the deep learning approaches, the stable architecture of our proposal limits the number of training parameters, thus restricting the model over-fitting. Moreover, the use of MODWT-based MRA transformation generates the wavelet and scale coefficients that can overlook the signal through noise resulting in accurate long-term forecasts. Hence, experimentally we can infer that the proposed XEWNet model can generate reliable point forecasts for dengue incidence cases in the considered geographical regions with appropriate lead time, more preferably for long-term forecasting.

\section{Significance of Improvements and Validation of Results} \label{stat_signif}
This section focuses on determining the robustness of the proposed XEWNet framework. We also describe the potential threats that impact the performance of our prediction system. It is important to note that the validation of the proposed model's performance depends on these factors to a considerable extent. 

\subsection{Statistical Significance of Performance}
\begin{figure}
    \centering
    \includegraphics[width=0.45\textwidth]{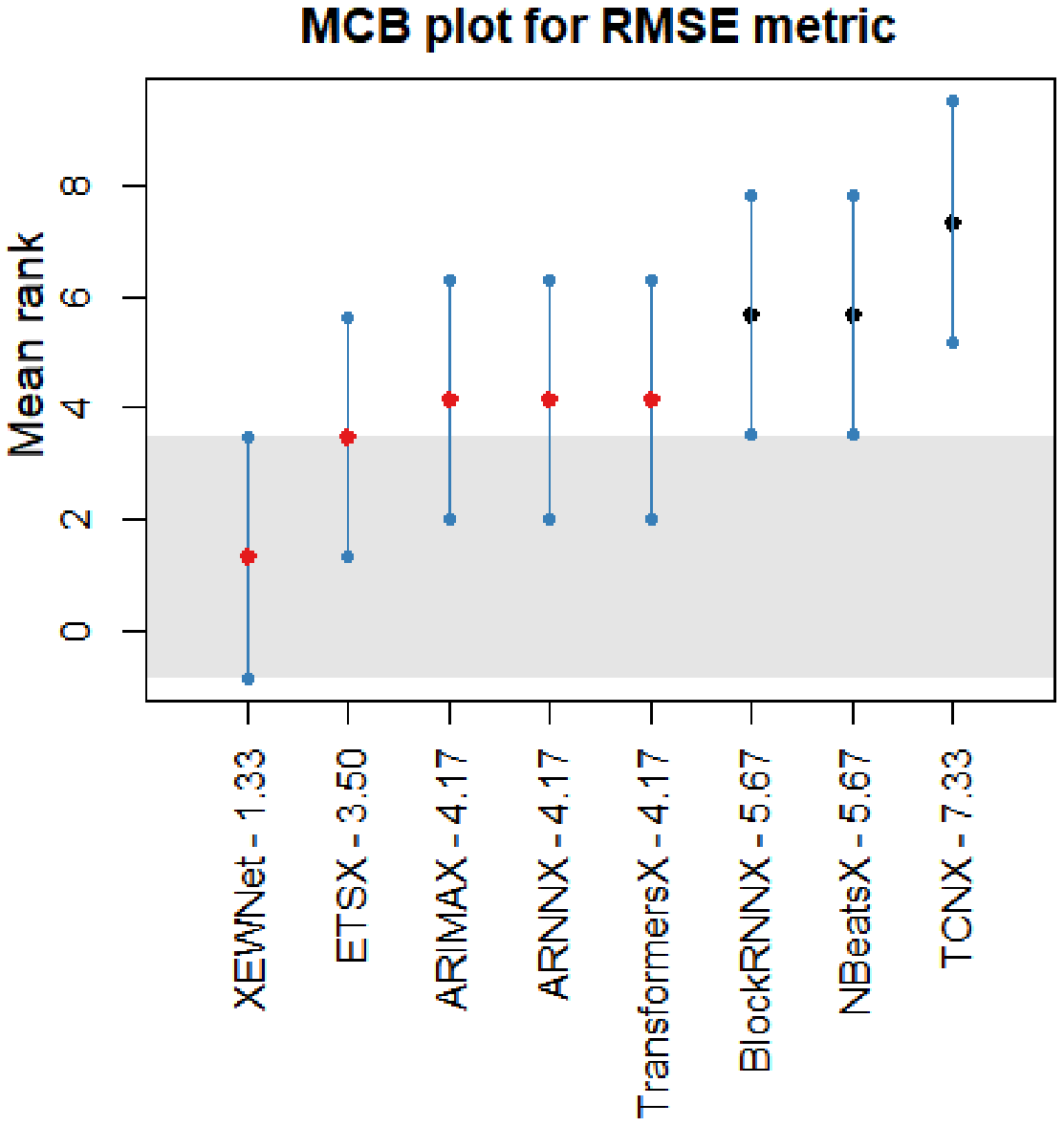}
    \includegraphics[width=0.45\textwidth]{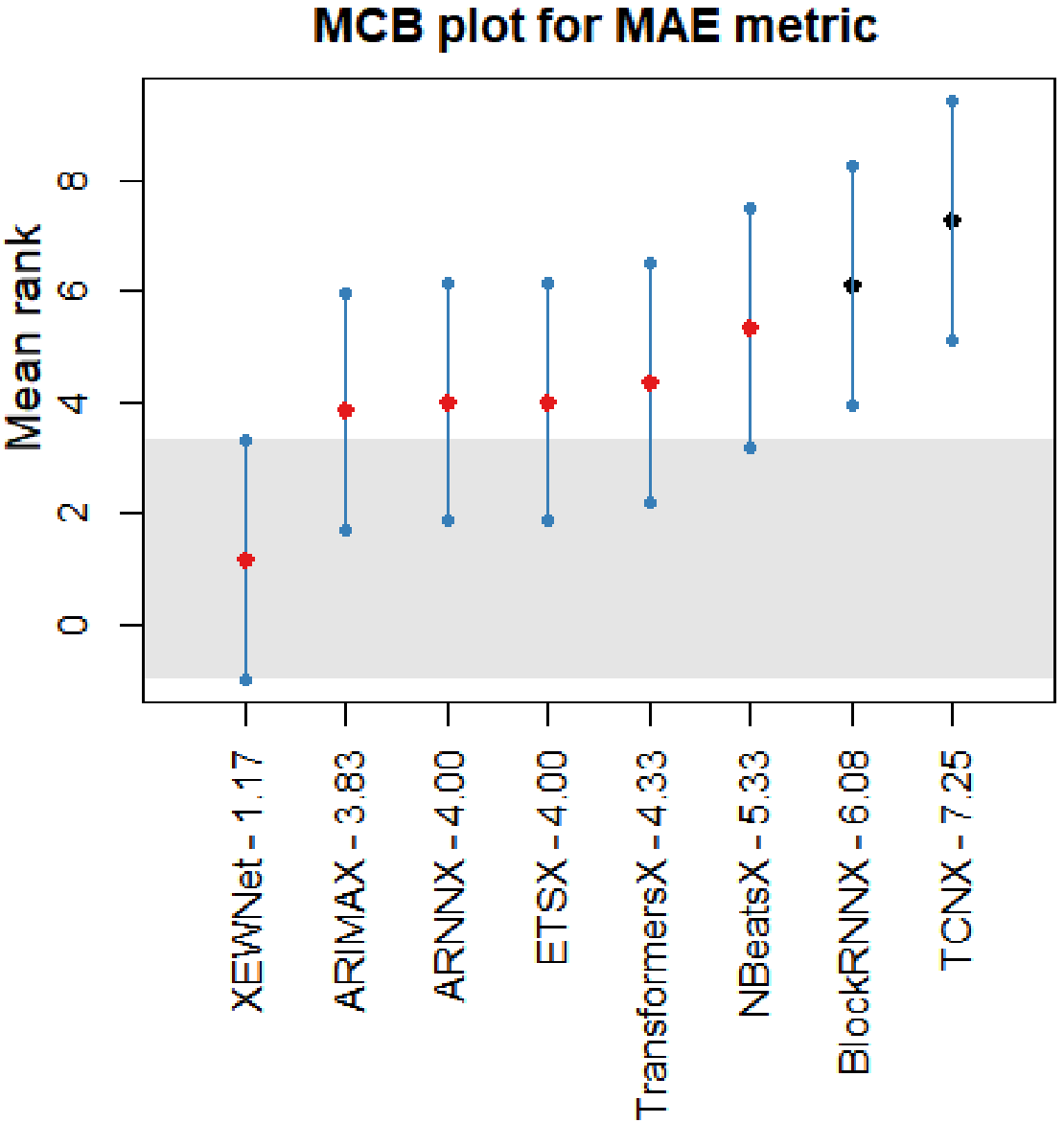}
    \caption{Schematic visualization of the multiple comparisons with the best (MCB) test. The left plot provides the result for the RMSE metric and the right plot indicates the MAE metric-based output. For example, in the figure, XEWNet-1.33 specifies the rank of the XEWNet model as 1.33 for the RMSE metric. The blue lines indicate the critical distance of the model, the middle point of this interval, which is denoted by black (significant) or red (not significant), represents the mean rank, and the shaded region marks the reference value.}
    \label{MCB_Plot}
\end{figure}

\begin{figure}
    \centering
    \includegraphics[width=0.48\textwidth]{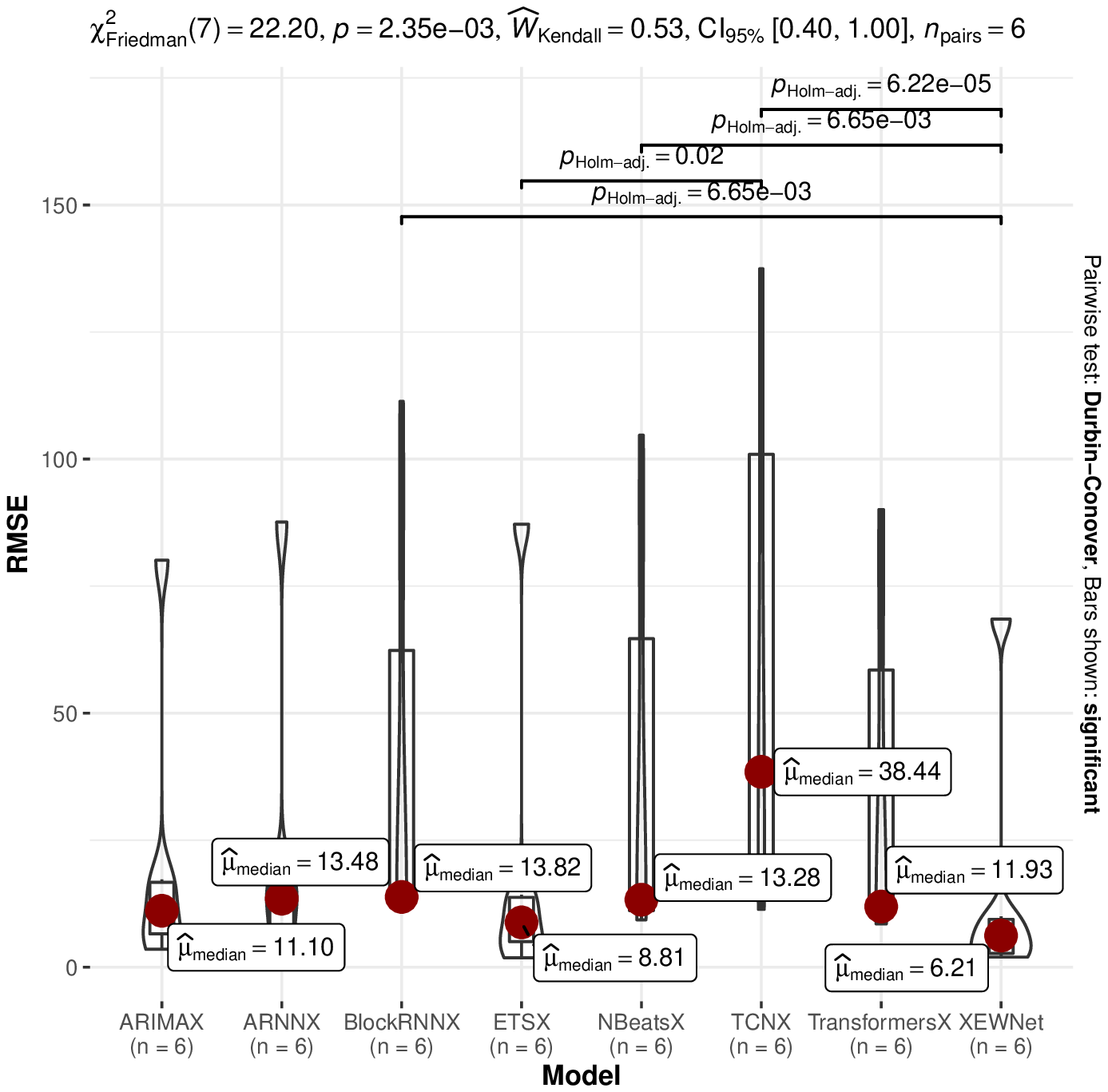}
    \includegraphics[width=0.48\textwidth]{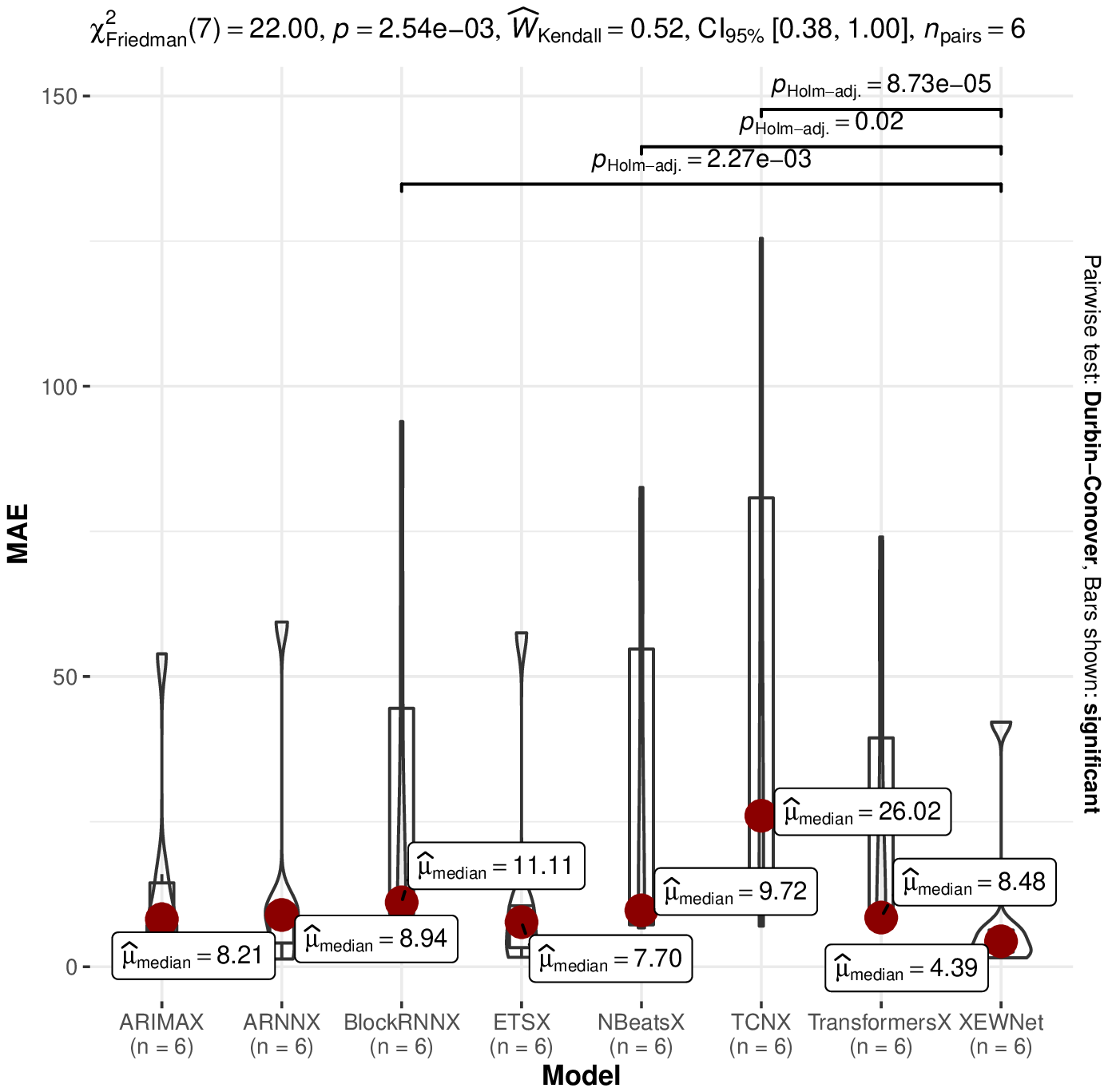}
    \caption{Friedman test and Durbin-Conover test results for the forecasting methods based on RMSE score (left) and MAE score (right). In the figure, the value of the Friedman test statistic is reported in the top panel with the confidence interval. The significant p-values for pairwise tests are provided along the horizontal bars. The median ranks of the forecasters are marked in red.}
    \label{RMSE_Friedman}
\end{figure}

In this section, we determine the statistical significance of the competitive forecasters with covariate (exogenous variable, i.e., rainfall) using several robust statistical tests. We restrict our attention to RMSE and MAE performance indicators for conducting these tests. Initially, we utilize the non-parametric multiple comparisons with the best (MCB) approach to compare the relative performance of different forecasting frameworks that include exogenous variables \cite{edwards1983multiple}. This statistical methodology assigns a rank to the competitive models based on their accuracy metric and computes their 95\% critical interval. The critical distance for the best performing forecaster serves as the reference value for this test. The results of the MCB test reported in Fig. \ref{MCB_Plot} indicate that the XEWNet model is the `best' among other forecasters with an average rank of 1.33 (for RMSE) and 1.17 (for MAE). From the RMSE plot (left), we notice that ETSX is the second best performing model (rank 3.50) and the average rank of ARIMAX, ARNNX, and TransformersX are precisely the same (rank 4.00). However, for the other deep learning models, there is a considerable increase in the RMSE score. Therefore, the critical interval of the proposed XEWNet model (shaded region) is considered the reference value. Since the critical distance of deep learning frameworks, namely NBeatsX, BlockRNNX, and TCNX, do not overlap with this reference value, their performance is significantly worse than the proposed XEWNet model. Moreover, for the MAE metric (right plot), we observe a considerable difference between the average rank of the proposed XEWNet model and the other forecasters. The non-overlapping nature of the reference value and the critical interval of BlockRNNX and TCNX leads to the conclusion that their performance is significantly worse than the best performing XEWNet model at a 95\% level of significance.
Alongside the MCB testing approach, we also consider the non-parametric Friedman test to validate our experimental results \cite{friedman1937use}. This distribution-free test is more suitable for non-normal datasets, a principal characteristic of real-world dengue incidence cases (refer to Table \ref{table_granger_test}). Furthermore, the Friedman test hypothesizes that the forecasters' performance is equivalent based on their average ranks. To conduct this statistical test, we compute the mean rank of the models w.r.t. their accuracy measures. 
This distribution-free test rejects the null hypothesis if the computed p-value is less than the chosen significance level. From the Friedman test results reported in Fig. \ref{RMSE_Friedman} w.r.t RMSE metric (left) and MAE metric (right), we obtain the computed p-value $<$ 0.05. Thus following the Friedman test procedure, we reject the null hypothesis of model equivalence and conclude that the performance of the forecasters is significantly different. Furthermore, we perform posthoc analysis using the Durbin-Conover test with a Holm correction for pairwise comparison \cite{conover1999practical}. This test evaluates the null hypothesis that there is no significant difference between the performance of each pair of forecasters using multiple comparisons. The non-parametric Durbin-Conover testing procedure rejects the null hypothesis if the computed p-value is less than the chosen significance level (0.05) and concludes that the model's performance is significantly different. We report the significant p-values of this pairwise test in Fig. \ref{RMSE_Friedman} for the RMSE metric (left) and MAE metric (right). The median rank of 6.21 (RMSE) and 4.39 (MAE) for the proposed XEWNet model are the least among all the forecasters. There is a significant difference in performance between XEWNet and BlockRNNX (p-value $< \; 0.05$), XEWNet and NBeatsX (p-value $< \; 0.05$), and XEWNet and TCNX (p-value $< \; 0.05$) in terms of RMSE metric. For the MAE score, the computed p-values between proposed XEWNet and TCNX, BlockRNNX, and NBeatsX are considerably lesser than $0.05$. Hence their performance varies significantly w.r.t. the proposal. Thus combining the results obtained from the non-parametric statistical tests, we can conclude that the point estimates of future dengue incidence generated by the proposed XEWNet framework are statistically validated and can be deployed by public-health officials as an early warning system in dengue-prone regions.

\subsection{Threats to Validity}
Finally, from the experimental evaluation (refer to Section \ref{Performance}), we observe that the proposed XEWNet outperforms all the benchmark forecasters in 75\% of the cases. In particular, the proposal generates the best long-term forecasts for all three geographical regions regarding the four accuracy measures. Our study utilizes RMSE, MAE, SMAPE, and MASE metrics as the key performance indicators based on their popularity in forecasting literature \cite{hyndman2018forecasting, chakraborty2020nowcasting}. Several other performance measures exist in the time series literature, and the appropriate metric choice can influence the forecasting performance. However, our experiments' performance indicators are carefully selected by including square, absolute, percentage, and scaled errors to ensure the reliability of the experimental results. In future work, other accuracy measures may need to be considered. Also, this study utilizes three publicly available dengue incidence cases and rainfall datasets to demonstrate the practical utility of our forecasting system. These time series datasets from San Juan, Iquitos, and Ahmedabad have been used widely in previous studies for dengue forecasting \cite{deb2022ensemble, johansson2019open, enduri2017estimation}. Our selected datasets are accumulated from various dengue endemic regions across the world and they are diverse in size and intensity of infection. For stronger generalizability of our findings, the forecasting performance of the proposed XEWNet for other affected regions may need to be further explored.

\section{Discussion}\label{discussion}
Dengue fever is a viral mosquito-borne disease transmitted predominately by the Aedes aegypti and Aedes albopictus mosquitoes. It infects an estimated 400 million people annually, with nearly half the world's population at risk of infection. Dengue cases are influenced by complex interactions of ecology, environment, meteorological factors, and virus factors, among many others. This data-driven study approaches the problem of forecasting dengue incidence (weekly cases for three regions) with the help of past dengue case data and rainfall data from the San Juan, Iquitos, and Ahmedabad regions. In addition, with the use of causality tests, we confirmed that incorporating weather-based dengue early warning into the dengue prediction system may significantly enhance the system's accuracy. 

In this paper, we proposed a new variant of the wavelet-based forecasting technique using the ARNN model, namely the XEWNet model. The proposed model is first theoretically built using a MODWT transformation applied to dengue incidence time series data and building locally auto-regressive neural networks on the decomposed time series. In the final stage, an ensemble of local forecasts is used to generate the out-of-sample forecasts of the XEWNet model. The final output of the model gives accurate and simultaneous estimates of the outbreak for the short-term (26 weeks) and long-term (52 weeks). This forecasting system will effectively implement invaluable control measures to control or even eliminate the cyclical dengue epidemic in the Latin American and Asian subcontinents considered in this study. For benchmark comparison, we have deployed several statistical, machine learning, and deep learning models using various statistical accuracy metrics. The results presented here indicate that XEWNet performs superior in most cases. Thus, the point estimates generated by our proposed XEWNet are more reliable and accurate for their practical usage in real-time dengue monitoring systems. However, since the proposed XEWNet model builds an ensemble of several auto-regressive neural nets, thus, there is a dependency between data sizes and model performances. When a lot of training data (dengue surveillance) is available and one wants to forecasts for the long-term (say, 52 weeks), our proposed framework seems more accurate and effective in generating reliable forecasts.
Further studies to improve the long-term sustainability of forecast precision and improvement of the XEWNet for generating probabilistic forecasts will help maintain a forecasting model's performance. Future research is also needed to address critical public health needs, produce probabilistic forecasts, and understand other major features that causally impact dengue incidence cases globally. An attempt to reveal and study the dengue mortality data and propose to study other key indicators such as peak intensity, epidemic duration, and rate of emergence of dengue epidemic can be considered as the future scope of this study.

\subsection*{Conflict of interest}
The authors declare that they have no conflict of interest.

\subsection*{Data availability statement}
All the codes and data are available in our GitHub repository:\\ \url{https://github.com/mad-stat/XEWNet}.

\bibliographystyle{plain}
\biboptions{square}
\bibliography{manuscript}

\section{Appendix}
\subsection{Discrete Wavelet Transform (DWT)} \label{DWT_formulation}
The Daubechies class of wavelets comprising orthogonal wavelets defines a discrete wavelet transform \cite{daubechies1992ten}. To formulate a DWT, we consider the Daubechies family of wavelets and denote the wavelet filters and the scaling filters as $\{a_m: m = 0,1,\ldots,\mathcal{M}-1\}$ and $\{b_m: m = 0,1,\ldots,\mathcal{M}-1\}$, respectively, where $\mathcal{M}-1$ is the length of the filter. The wavelet filters and scaling filters are restricted to satisfy the orthonormality property:
\begin{equation}
    \sum_{m=0}^{\mathcal{M}-1}a_m^2 = \sum_{m=0}^{\mathcal{M}-1}b_m^2 = 1, %\;\;\; \text{Unit energy assumptions}
    \label{eq_prop_1}
\end{equation}
\begin{equation}
    \sum_{m=0}^{\mathcal{M}-1}a_m a_{m + 2n} =\sum_{m=0}^{\mathcal{M}-1}b_m b_{m + 2n} = 0   \; ; \; \forall \; \;  n \in \mathbf{Z}_{\ne 0}. % \;\;\; \text{Even-length scaling assumptions}
    \label{eq_prop_2}
\end{equation}
In the Wavelet literature, Eqn. \ref{eq_prop_1} is referred as unit energy assumptions and Eqn. \ref{eq_prop_2} as even-length scaling assumptions \cite{percival2000wavelet}. Furthermore, the wavelet filters $\{a_m\}$ and the scaling filters $\{b_m\}$ are also linked by the following constraint(s):
\begin{equation*}
    b_m \equiv (-1)^{m+1} a_{\mathcal{M}-1-m} \;\;\;\;
    \textrm{ or } \;\;\;\; a_m \equiv (-1)^m b_{\mathcal{M}-1-m}; \; \; 
    \text{for} \; \; m = 0, 1, \ldots, \mathcal{M}-1.
\end{equation*}
Thus, scaling filters are referred to as the ``quadrature mirror" filter corresponding to wavelet counterparts. The above-stated formulation of the DWT coefficients is popularly known as the ``pyramid algorithm" \cite{percival1997analysis}. Suppose we represent the time series as $Y = \{Y_t: t= 1,2 \ldots, N\}$. Considering the $l^{th}$ stage input to the pyramid algorithm as $\{W_{l-1,t}:t=0,\ldots,N_{l-1}-1\}$, where $N_l = {N}/{2^l}$ and $W_{0,t} \equiv Y_t$, then following \cite{walden2001wavelet}, the corresponding outputs of the algorithm are the $l^{th}$ level scaling and wavelet coefficients of the series $\{Y_t\}$ denoted by
\begin{equation*}
    U_{l,t} = \sum_{m=0}^{\mathcal{M}_l - 1} a_{l,m} Y_{(2^l(t+1)-1-m)\text{ mod }N} \;\;\;\; \text{ and } \;\;\;\;
    W_{l,t} = \sum_{m=0}^{\mathcal{M}_l - 1} b_{l,m} Y_{(2^l(t+1)-1-m)\text{ mod }N},
\end{equation*}
where $\mathcal{M}_l = (2^l -1)(\mathcal{M}-1)+1$. These $l^{th}$ level scaling and wavelet filters satisfy the orthonormality properties as stated in Eqn. (\ref{eq_prop_1}) and Eqn. (\ref{eq_prop_2}) alongside
\begin{equation*}
    \sum_{m=0}^{\mathcal{M}_l-1} b_{l,m} = 2^{{l}/{2}}  \textrm{ and } \sum_{m=0}^{\mathcal{M}_l-1} a_{l,m} = 0.
\end{equation*}
The pyramid algorithm formulation of the DWT has been applied in several domains \cite{nazaripouya2016univariate,chen2017high} for the efficient extraction of signal from noise. 

\subsection{Multi-resolution analysis of MODWT algorithm}\label{MODWT_MRA_Mathematical}
The maximal overlap discrete wavelet transform (MODWT) is a modified version of DWT in that it is a non-orthogonal and redundant transformation \cite{percival2000wavelet}. The wavelet filters $\{\tilde{a}_m\}$ and the scaling filters $\{\tilde{b}_m\}$ of MODWT can be defined as normalized versions of DWT filters as follows:
\begin{equation}\label{eq:3}
    \tilde{a}_{l,m} = \frac{a_{l,m}}{2^{l/2}} \;\;\;\; \textrm{ and } \;\;\;\; \tilde{b}_{l,m} = \frac{b_{l,m}}{2^{l/2}}.
\end{equation}
Following \cite{percival2000wavelet}, the MODWT wavelet and scaling coefficients of level $l$, obtained as the $l^{th}$ stage output of the pyramid algorithm, are given by:
\begin{equation}\label{MODWT_pyramid_eqn}
  \tilde{U}_{l,t} = \sum_{m=0}^{\mathcal{M}_l-1}\tilde{a}_{l,m} Y_{(t-m) \textrm{ mod }N} \;\;\;\; \textrm{ and } \;\;\;\;
    \tilde{W}_{l,t} = \sum_{m=0}^{\mathcal{M}_l-1}\tilde{b}_{l,m} Y_{(t-m) \textrm{ mod }N},
\end{equation}
where $\mathcal{M}_l = (2^l-1)(\mathcal{M}-1)+1$. Thus, the MODWT coefficients at level $l$ are defined as the convolutions of the original time series $\{Y\}$, and it has the same length as that of $\{Y\}$. Meanwhile, Eqn. \ref{MODWT_pyramid_eqn} can be expressed in matrix notation as:
\begin{equation*}
  \tilde{U}_l = \overline{\overline{{u}}}_l Y \;\;\;\; \textrm{ and } \;\;\;\; \tilde{W}_l = \overline{\overline{{w}}}_l Y,
\end{equation*}
where the $N \times N$ matrices $\overline{\overline{{u}}}_l$ and $\overline{\overline{{w}}}_l$ comprises of the wavelet coefficients $\{\tilde{a}_{l,m}\}$ and scaling coefficients $\{\tilde{b}_{l,m}\}$, respectively. Thus, the original signal $\{Y\}$ can be represented as:

\begin{equation*}
    Y = \sum_{l=1}^L \overline{\overline{{u}}}_l^T \tilde{U}_l + \overline{\overline{{w}}}_L^T \tilde{W}_L = \sum_{l=1}^L D_l + S_L,
\end{equation*}
where the details coefficient $D_l = \overline{\overline{{u}}}_l^T \tilde{U}_l$ denotes the irregular fluctuations of the series $\{Y\}$ at scale $l \; \forall \; (l = 1,2,\ldots,L)$ and the smooth $S_L = \overline{\overline{{w}}}_L^T \tilde{W}_L$ indicates the overall trend of the original signal at scale $L$ \cite{percival1997analysis}. The above-stated multi-resolution analysis (MRA) of MODWT utilizes a zero-phase filter to partition a signal across details and smooth coefficients. This scale-based additive decomposition generates good time resolution and poor frequency resolution for high-frequency signals and vice-versa for low-frequency signals.

\end{document}